\documentclass[preprint]{ptephy_v1}

\preprintnumber{YITP-24-36} 
\usepackage{hyperref}

\usepackage{hyperref}     
\usepackage{amsmath}       
\usepackage{amssymb} 
\usepackage{graphicx}
\usepackage{url}
\usepackage{enumerate}
\usepackage{color}
\usepackage{ulem}
 
\usepackage{float,wrapfig,color}

\def\eq#1{(\ref{#1})}
\def\s[#1\s]{\begin{align}\begin{split}#1\end{split}\end{align}}
\def\[#1\]{\begin{align}#1\end{align}}

\def\rhos{{\rho_{\rm signed}}}

\def\bpsi{\bar\psi}
\def\bvph{\bar\varphi}
\def\vph{\varphi}

\begin{document}

\title{Signed eigenvalue/vector distribution \\
of complex order-three random tensor}


\author{Naoki Sasakura}
\affil{Yukawa Institute for Theoretical Physics, Kyoto University, \\
and \\
CGPQI, Yukawa Institute for Theoretical Physics, Kyoto University, \\
Kitashirakawa, Sakyo-ku, Kyoto 606-8502, Japan
 \email{sasakura@yukawa.kyoto-u.ac.jp}}

\begin{abstract}%
We compute the signed distribution of the eigenvalues/vectors of the complex order-three random tensor
by computing a partition function of a four-fermi theory, where signs are from a Hessian determinant associated 
to each eigenvector.
The issue of the presence of a continuous degeneracy of the eigenvectors is properly treated by a gauge-fixing. 
The final expression is compactly represented by a generating function, which has an expansion whose powers are
the dimensions of the tensor index spaces. 
A crosscheck is performed by Monte Carlo simulations. 
By taking the large-\!$N$ limit we obtain a critical point where the behavior of the signed distribution qualitatively changes, 
and also the end of the signed distribution. The expected agreement of the end of the signed distribution 
with that of the genuine distribution provides a few applications, such as 
the largest eigenvalue, the geometric measure of entanglement, and the best rank-one approximation in the large-\!$N$ limit.
\end{abstract}

\subjectindex{A13, A45, B83, B86}

\maketitle

\section{Introduction}
\label{sec:introduction}
In random matrix theories, eigenvalue distributions play important roles. Wigner modeled atomic Hamiltonians
by random matrices, and derived the celebrated semi-circle law of eigenvalue distributions \cite{Wigner}. Eigenvalue
distributions are used in the techniques of solving random matrix models \cite{Brezin:1977sv,matrix}. 
Topological changes of eigenvalue distributions give intriguing insights into QCD dynamics 
\cite{Gross:1980he,Wadia:1980cp}.  

The tensor eigenvalues/vectors \cite{Qi,lim,cart,qibook} also
appear in various contexts, such as in quantum \cite{Biggs:2023mfn} and classical \cite{Evnin:2021buq} gravity, 
spin glasses \cite{pspin,pedestrians,randommat},  
quantum information theory \cite{shi,barnum,estimate}, 
computer science \cite{SAPM:SAPM192761164,Carroll1970,bestrankone, comon, spiked}, 
and others \cite{qibook},
though the terminologies, tensor eigenvalue/vector, are not always used. 
It is also expected that tensor eigenvalues/vectors are useful for the understanding of the dynamics of random 
tensor models \cite{Ambjorn:1990ge,Sasakura:1990fs,Godfrey:1990dt,Gurau:2009tw,Gurau:2024nzv}.
In these applications one may be interested in not only specific tensor cases but also 
general properties over arbitrary tensors. 
For the latter interest, various results have already been obtained for the real symmetric random tensor, 
such as the eigenvalue distribution \cite{randommat,secondmoment},
the expected number of the real eigenvalues \cite{fyodorov1,realnum1,realnum2}, 
an estimation of the largest eigenvalue \cite{Evnin:2020ddw},  
and an extension of the Wigner semi-circle law \cite{Gurau:2020ehg}. 

The results above for the real symmetric random tensor should be extended to other kinds of tensors.
For example applications to quantum information theory require considering complex 
tensors, which are generally not symmetric\footnote{For instance, a state $|\Psi\rangle$ in a product Hilbert space
${\cal H}_1\otimes {\cal H}_2 \otimes {\cal H}_3$ can be represented by a complex tensor as $|\Psi\rangle=C_{abc} | a \rangle_1| b \rangle_2| c \rangle_3$.}.
The complex eigenvalue/vector problem cannot just be solved by the analytic continuation of the real problem,
but rather there exist some classes which must separately be treated \cite{Qi,lim,cart,qibook}.
A version of the complex eigenvalue distribution of the symmetric complex random tensor was studied in 
\cite{Kent-Dobias:2020egr}, but there remain some other versions not yet studied.
Moreover, for non-symmetric tensors, the eigenvalue/vector problem is not represented by 
one equation but rather by a system of eigenvalue/vector equations \cite{qibook}. 
Therefore it would be interesting and challenging to compute the eigenvalue/vector distribution of 
the complex random tensor, which is generally not symmetric.
In Section~\ref{sec:app} 
we show an application of our result to a geometric measure of entanglement in quantum information theory  
(A closely related result was reported on the same day by another group \cite{Dartois:2024zuc}). 

Our strategy of computing tensor eigenvalue/vector distributions is to apply quantum field theoretical methods,
which was indeed successful for the case of the real symmetric random tensor \cite{Sasakura:2022zwc,Sasakura:2022iqd,Sasakura:2022axo,Sasakura:2023crd,Kloos:2024hvy}. 
The advantages of the approach are that it is in principle
systematically applicable to compute a wide range of statistical quantities of random tensors, and also 
that various sophisticated quantum field theoretical techniques, knowledge and insights can be used.
As far as random tensors are Gaussian, various distributions can be reduced to
computations of partition functions of certain zero-dimensional quantum field theories with four-interactions. 
The easiest is the computation of signed distributions 
of eigenvalues/vectors, where signs come from the determinant of a certain Jacobian matrix which is associated to each 
eigenvalue/vector (details appear in Section~\ref{sec:strategy}). 
The quantum field theories for such signed distributions are four-fermi theories, and in principle the partition functions can 
exactly be computed as polynomial functions of couplings\footnote{Except that overall factors contain
exponentials of inverse couplings.}, 
since fermion integrations \cite{zinn} are similar to taking derivatives.  

While a signed distribution is generally different from the corresponding genuine one, 
they may coincide near an end of the distributions in the large-\!$N$ limit. 
This is indeed true for the signed and the genuine eigenvalue/vector distributions of the real symmetric 
random tensor \cite{parisi,example,randommat}, as explicitly demonstrated in detail in \cite{Kloos:2024hvy}.
As explained in Section~\ref{sec:strategy}, this coincidence may generally be expected, when an
eigenvalue/vector problem comes from an optimization problem. Therefore it may be possible
to obtain the location of the end of a genuine distribution from the corresponding signed distribution. 
Since the end of an eigenvalue/vector distribution gives the typical optimized value of an optimization problem, 
a signed distribution is not only the easiest to compute but also very useful for applications \cite{Kloos:2024hvy}. 

In this paper we compute the signed distribution of the eigenvalues/vectors  of the complex order-three random tensor
by computing the partition function of a four-fermi theory.
There is an issue that the eigenvectors have a continuous degeneracy, which
prevents us from a naive application of the method previously taken in
\cite{Sasakura:2022zwc,Sasakura:2022iqd,Sasakura:2022axo,Sasakura:2023crd,Kloos:2024hvy}. 
To properly deal with this issue, we embed gauge fixing conditions to the eigenvector equations, and  
extract the gauge independent distribution at the final stage.
The final expression can compactly be expressed by a generating function, which has an expansion whose powers are 
the dimensions of the index spaces. 
By taking the large-\!$N$ limit, we obtain a critical point where the behavior of the signed distribution qualitatively changes 
and also the end of the signed distribution. We notice that a method using the Schwinger-Dyson equation is efficient
for the computations of these quatities.
The overall character is very similar to the case of the real symmetric random tensor \cite{Kloos:2024hvy}, 
strongly suggesting that the signed and the genuine distributions coincide in the vicinities of the ends. Assuming this, 
we explain a few applications, the largest eigenvalue, the geometric measure of entanglement (See also \cite{Dartois:2024zuc}),
and the best rank-one approximation in the large-\!$N$ limit. 

This paper is organized as follows. 
In Section~\ref{sec:egeqs}, we discuss the system of eigenvector equations for the 
complex order-three random tensor. The issue of the presence of a continuous degeneracy of solutions
is properly treated by embedding gauge fixing conditions to the eigenvector equations.
In Section~\ref{sec:strategy}, we explain our procedure of expressing 
a tensor eigenvector distribution as a partition function of a zero-dimensional quantum field theory with four-interactions. 
In particular a signed distribution can be rewritten as a four-fermi theory.
In Section~\ref{sec:fourfermi}, we obtain the four-fermi theory corresponding to the present case.
In Section~\ref{sec:compfermi}, we obtain the signed eigenvector distribution by 
computing the partition function of the four-fermi theory. It can compactly 
be expressed by a generating function, which has an expansion whose powers are the dimensions of the index spaces.
In Section~\ref{sec:gaugeinv}, we extract the gauge independent signed distribution. We point out
its topological property. We perform Monte Carlo simulations for crosscheck.
In Section~\ref{sec:largen}, we take the large-\!$N$ limit by a method with the Schwinger-Dyson equation
in Section~\ref{sec:sd}, and by a saddle point analysis in Section~\ref{sec:largenfinal}.
We find a critical point where the behavior of the signed distribution qualitatively changes,
and also find the end of the distribution.
In Section~\ref{sec:app}, we explain a few applications of the location of the end, such as 
the largest eigenvalue, the geometric measure of entanglement,
and the best rank-one approximation in the large-\!$N$ limit.
Section~\ref{sec:summary} is devoted to a summary and future prospects.
 
\section{Eigenvector equations for complex tensors and gauge fixing}
\label{sec:egeqs}
In this paper the tensor is complex and has only three independent indices for simplicity.
The dimensions of the index spaces are denoted by $N_i\ (i=1,2,3)$. Namely, tensor $C$ has components, 
\[
C_{a_1 a_2 a_3} \in \mathbb{C}, \ (a_1=1,2,\ldots, N_1,\ a_2=1,2,\ldots, N_2,\  a_3=1,2,\ldots, N_3).
\] 
The eigenvector equation we consider is a system of equations \cite{qibook},
\s[
&C_{a_1a_2a_3} v_{(2)}^{*a_2} v_{(3)}^{*a_3}=v_{(1) a_1} ,\\
&C_{a_1a_2a_3} v_{(3)}^{*a_3} v_{(1)}^{*a_1}=v_{(2) a_2} ,\\
&C_{a_1a_2a_3} v_{(1)}^{*a_1} v_{(2)}^{*a_2}=v_{(3) a_3} .
\label{eq:egeqs}
\s] 
Here contracted upper and lower indices are assumed to be summed over throughout this paper, unless otherwise stated.
The symbol $*$ denotes taking the complex conjugation. $v_{(i)}\ (i=1,2,3)$ denote vectors in 
the $i$-th index space. The system \eq{eq:egeqs} is a natural non-symmetric generalization of the eigenvector equation 
for symmetric tensors\footnote{The eigenvector equation for symmetric tensors is  $C_{abc}v_b v_c =v_a$ 
in \cite{Sasakura:2022zwc,Sasakura:2022iqd,Sasakura:2022axo,Sasakura:2023crd,Kloos:2024hvy} .}. 

The system $\eq{eq:egeqs}$ is invariant under unitary transformations.
The lower and upper index spaces transform under the unitary transformations and their complex conjugates, respectively.
Namely, the unitary transformations are given by
\s[
&C’_{a_1a_2a_3}={U^{(1)}_{a_1}}^{a_1'}{U^{(2)}_{a_2}}^{a_2'}{U^{(3)}_{a_3}}^{a_3'} C_{a_1'a_2'a_3'}, \\
&v’_{(i) a}=U^{(i)}_a{}^{a'} v_{(i) a'}, 
\label{eq:unitary}
\s] 
where $U^{(i)} \ (i=1,2,3)$ are the $N_i$-dimensional unitary transformations in the fundamental representations, 
which satisfy $\sum_{a=1}^{N_i} U^{(i)}_{a}{}^{a'}U^{(i)a''*}_a=\delta_{a''}^{a'}$.

An arbitrarily given tensor $C$ generally breaks the unitary symmetry above 
except for the following two-dimensional Abelian symmetry:
\s[
U^{(i)}_{a}{}^{a'}=e^{I \theta_i} \delta_{a}^{a'}, \hbox{ with } \sum_{i=1}^3 \theta_i=0,
\label{eq:2dimu1}
\s]
where $I$ denotes the imaginary unit, and $\theta_i$ are real. 
Therefore the solutions to the eigenvector equation \eq{eq:egeqs} have the following continuous degeneracy, 
\[
v_{(i)a}'=e^{I \theta_i} v_{(i)a}, \hbox{ with } \sum_{i=1}^3 \theta_i=0.
\label{eq:rotation}
\] 
In this paper we remove this degeneracy of the solutions for the number counting of the solutions,
namely, we count one for each gauge orbit \eq{eq:rotation}. 
There are various ways of fixing this degeneracy. We impose the following two
gauge fixing conditions on the solutions:
\s[
v_{(i)} \cdot v_{(i)} \ (i=2,3) \hbox{ are real},
\label{eq:gaugefix}
\s]
where $v_{(i)} \cdot v_{(i)}=\sum_{a=1}^{N_i}  v_{(i)a} v_{(i)a}$\footnote{Note that the symbol $\cdot$
also allows summing over two upper/lower indices. This breaks the unitarity symmetry, which is needed for a gauge fixing.}. 
It is obvious that any solutions can be transformed to satisfy \eq{eq:gaugefix} by
applying \eq{eq:rotation}. Note that the gauge fixing condition \eq{eq:gaugefix} is invariant under 
residual discrete unitary symmetries: a solution still has a 16-fold  degeneracy given by
\[
\pm v_{(i)} ,\ \pm I v_{(i)}, (i=2,3),
\label{eq:residual}
\]
accompanied with appropriate transforms of $v_{(1)}$.
The degeneracy will be removed in Section~\ref{sec:gaugeinv} by appropriately interpreting the probability density. 

The procedure taken in the previous 
papers \cite{Sasakura:2022zwc,Sasakura:2022iqd,Sasakura:2022axo,Sasakura:2023crd,Kloos:2024hvy} 
to rewrite the eigenvector distributions as quantum
field theories works only when the number of independent equations is the same as that of variables\footnote{Otherwise, 
the determinant of the Jacobian matrix and the delta functions, which appear later, will vanish and diverge, respectively.
The product could be finite but seems difficult to properly handle.}.
In the present case, because of the two-dimensional degeneracy under the symmetry \eq{eq:rotation} 
the system \eq{eq:egeqs} has two less independent equations than the number of variables. In fact, 
by contracting each of \eq{eq:egeqs} with $v_{(i)}^*\ (i=1,2,3)$, we obtain the same equation 
Im($C_{a_1a_2a_3} v_{(1)}^{*a_1} v_{(2)}^{*a_2} v_{(3)}^{*a_3})=0$ from three different equations.

A useful fact is that the shortage of the number of independent equations is the same as that of the gauge fixing 
conditions \eq{eq:gaugefix}. 
Therefore one may modify the equations \eq{eq:egeqs} so that the gauge fixing conditions \eq{eq:gaugefix}
are included. 
A natural manner for the present case is to consider the set of equations $f^{(i)}_a=0$ with  
\s[
&f^{(1)}_{a_1}=-C_{a_1a_2a_3} v_{(2)}^{*a_2} v_{(3)}^{*a_3}+v_{(1) a_1},\\
&f^{(2)}_{a_2}=-C_{a_1a_2a_3} v_{(3)}^{*a_3} v_{(1)}^{*a_1}+v_{(2) a_2}\left( 1+ \beta 
\left( v_{(2)} \cdot v_{(2)} - v^*_{(2)} \cdot v^*_{(2)} \right)\right) ,\\
&f^{(3)}_{a_3}=-C_{a_1a_2a_3} v_{(1)}^{*a_1} v_{(2)}^{*a_2}+v_{(3) a_3}\left(1+
\beta \left( v_{(3)} \cdot v_{(3)} - v^*_{(3)} \cdot v^*_{(3)} \right)\right),
\label{eq:newsystem}
\s] 
with an arbitrary real parameter $\beta$.
In fact, by contracting each of \eq{eq:egeqs} with $v_{(i)}^*\ (i=1,2,3)$, one finds
\[
|v_{(1)} |^2 =| v_{(2)} |^2\left(1+\beta
\left( v_{(2)} \cdot v_{(2)} - v^*_{(2)} \cdot v^*_{(2)} \right) \right)
=| v_{(3)} |^2\left(1+\beta
\left( v_{(3)} \cdot v_{(3)} - v^*_{(3)} \cdot v^*_{(3)} \right) \right),
\label{eq:contracteqs}
\]
where $| v|$ denotes the norm of a vector $v$, namely, $|v|=\sqrt{v_a v^{*a}}$.
Since the terms multiplied by $\beta$ are imaginary, \eq{eq:contracteqs} implies
\s[
&| v_{(1)} |^2 =|v_{(2)} |^2 =| v_{(3)} |^2 ,\\
&v_{(2)} \cdot v_{(2)} - v^*_{(2)} \cdot v^*_{(2)}=0,\\
&v_{(3)} \cdot v_{(3)} - v^*_{(3)} \cdot v^*_{(3)}=0,
\label{eq:veq}
\s]
where the latter two are the gauge fixing conditions \eq{eq:gaugefix}, as demanded.
Since the additional terms multiplied by $\beta$ 
in the new system \eq{eq:newsystem} vanishes under the gauge fixing conditions,
solving the new system \eq{eq:newsystem} is equivalent to solving \eq{eq:egeqs} with the gauge fixing 
conditions. This is the setup we wanted, and we will therefore analyze \eq{eq:newsystem} in the rest of this paper.

\section{Quantum field theory expressions of distributions}
\label{sec:strategy}
Let us explain our strategy taken in the previous papers
\cite{Sasakura:2022zwc,Sasakura:2022iqd,Sasakura:2022axo,Sasakura:2023crd,Kloos:2024hvy} 
to compute the eigenvalue/vector distributions of tensors.
In fact this strategy is widely applicable to various problems, so we will explain it in general forms in this section.
Suppose that there is a set of real equations $f_i(x,C)=0 \ (i=1,2,\cdots, \bar N)$, where $x$ represents the real variables 
$x_i\ (i=1,2,\cdots,\bar N)$ to be solved, and $C$ represents random real variables $C_j\ (j=1,2,\ldots,\# C)$.
It is also assumed that $f_i(x,C)$ are linear in $C$, while they can generally be non-linear in $x$. 
In the present case of Section~\ref{sec:egeqs}, the equations are complex for complex variables, but they can be 
separated into the real and imaginary parts to make them all real. 
The probability distribution of the solutions $x$ for a given $C$ under the measure 
$\prod_{i=1}^{\bar N} dx_i$ is given by
\[
\rho(x,C)=\sum_{k=1}^{\# {\rm sol}(C)} \prod_{i=1}^{\bar N} \delta\left( x_i-x^{(k)}_i(C) \right),
\label{eq:deltasol}
\]
where $x^{(k)}(C)\ (k=1,2,\ldots, \# {\rm sol}(C))$ denote all the solutions\footnote{Trivial solutions like $x=0$
may be ignored, depending on problems.} to the set of equations, $f_i(x,C)=0$. 

When $C$ has a probability distribution, the mean of the distribution of $x$ is given by 
\[
\rho(x)=\langle \rho(x,C) \rangle_C,
\label{eq:genuine}
\]
where $\langle \cdot \rangle_C$ denotes taking the expectation value under the probability distribution of $C$. 
 
Since $x^{(k)}(C)$ are the solutions to $f_i(x,C)=0$, \eq{eq:deltasol} can be 
rewritten as
\[
\rho(x,C)=\left | \det M(x,C) \right | \prod_{i=1}^{\bar N} \delta\left( f_i(x,C)\right), 
\label{eq:rhoabs}
\]
where $\left |{\rm det} M(x,C) \right |$ denotes the absolute value of the determinant of the Jacobian,
\[
M(x,C)_{ij} =\frac{\partial f_j(x,C)}{\partial x_i},\ (i,j=1,2,\ldots,\bar N).
\]

The Jacobian factor with the absolute value in \eq{eq:rhoabs} is treatable by the procedure employed in \cite{Sasakura:2022axo}, where
bosons and fermions are introduced, or by a replica trick applied to a fermionic theory \cite{Sasakura:2022iqd}. 
On the other hand, we can also consider a distribution without taking the
absolute value, as previously employed in \cite{Sasakura:2022zwc}. We call it a signed distribution, because 
the expression corresponding to \eq{eq:deltasol} has an extra sign:
\s[
\rho_{\rm signed}(x,C)&:=\det M(x,C) \prod_{i=1}^{\bar N} \delta\left( f_i(x,C)\right) \\
&= \sum_{k=1}^{\# {\rm sol}(C)} {\rm Sign}\left( \det M(x^{(k)},C)\right)  
\prod_{i=1}^{\bar N} \delta\left( x_i-x^{(k)}_i(C) \right).
\label{eq:rhosign}
\s]
The signed distribution of $x$ is defined by
\[
\rho_{\rm signed} (x)=\left\langle \rho_{\rm signed} (x,C) \right \rangle_C.
\label{eq:rhosign2}
\]

There are a few reasons to make this signed distribution important, though it is different from the genuine distribution. 
One is the technical reason that the signed distribution is much easier to compute.
Without taking the absolute value, the determinant factor can straightforwardly be represented by using a formula 
$\det M =\int d\bar \psi d\psi\, e^{\bar \psi M \psi}$ \cite{zinn} with a couple of fermions $\bar\psi,\psi$.
Then, as we will see below, the signed distribution is expressed as a partition function 
of a four-fermi theory. Since fermion integrals \cite{zinn} are essentially taking derivatives,
the partition function can in principle exactly be computed as a polynomial function of couplings for any finite $N$. 
 
Another reason is that a signed distribution may coincide with a genuine distribution 
up to an overall sign factor in the vicinities of their ends in the large-\!$N$ limit, 
though they are generally different in the other regions. 
This can generally be expected in the case that an eigenvalue/vector equation is a stationary 
point equation of a potential of an optimization problem.
In such a case, the matrix $M(v,C)$ is a Hessian on each stationary point of a potential, and it takes 
a common signature, all positive (or negative), over most of the stationary points 
in the vicinity of an end of a distribution, since 
most of the stationary points near an end are stable (or maximally unstable). In particular
the ends of a signed and a genuine distributions should agree. 
Since such an end represents the typical optimized value of an optimization problem, 
a signed distribution can be used to compute it, though it is different from 
the genuine distribution.

By considering the case that $C$ obeys the Gaussian distribution with mean value zero, introducing fermions for the 
determinant, and using a standard formula $\int_{\mathbb{R}}d\lambda \, e^{I \lambda y}=2 \pi\delta(y)$, one can 
rewrite \eq{eq:rhosign2} as
\[
\rho_{\rm signed} (x)=A^{-1} (2 \pi)^{-\bar N} \int_{\mathbb{R}^{\# C}} dC \int_{\mathbb{R}^{\bar N}} d\lambda 
\int d\bar\psi d\psi  \, e^{S_{\rm bare}},
\label{eq:rerho}
\]
where $A= \int_{\mathbb{R}^{\# C}} dC e^{-\alpha C_i C_i }$,  $dC=\prod_{i=1}^{\# C} dC_i$,
$d\lambda=\prod_{i=1}^{\bar N} d\lambda_i$,
\[
S_{\rm bare}=-\alpha C_i C_i + \bar\psi_i M(x,C)_{ij} \psi_j + I \lambda_i f_i(x,C),
\label{eq:res}
\]
where $\alpha$ is a positive parameter, and repeated indices are assumed to be summed over. 
Since $C$ appears at most quadratic in $S_{\rm bare}$ 
from the assumption that $f$ are linear in $C$, the integral over $C$ and $\lambda$ can explicitly be performed
by Gaussian integrations.
Then we obtain a theory of a couple of fermions with four-fermi interactions, which is often called a four-fermi theory.

\section{The four-fermi theory}
\label{sec:fourfermi}
In this section 
we will apply the general procedure of Section~\ref{sec:strategy} to our present case of Section~\ref{sec:egeqs}, and 
explicitly obtain the corresponding four-fermi theory. 
We will use complex notations for convenience, differently from Section~\ref{sec:strategy}. 
Details of the complex conventions are given in Appendix~\ref{app:conv}. 

It is rather straightforward to derive the expression corresponding to \eq{eq:rerho} and \eq{eq:res} except for 
the following useful matter.  Since the determinant factor is multiplied by the delta functions of 
the eigenvector/gauge-fixing equations \eq{eq:newsystem} in \eq{eq:rhosign}, 
it is allowed to use the eigenvector equations and the gauge conditions to simplify the determinant factor. 
For example, a component of the matrix $M(v,C)$ has a simplified expression, 
\[
\frac{\partial  f^{(2)}_{a}}{\partial v_{(2) a'}}=\delta_{a}^{a'} +2 \beta v_{(2)a} v_{(2) a'}
\]
by using $v_{(2)} \cdot v_{(2)} - v^*_{(2)} \cdot v^*_{(2)}=0$ in \eq{eq:veq}.

By explicitly computing the matrix $M$, one can derive the expression corresponding to \eq{eq:rerho} as
\[
\rhos(v)=\frac{1}{A\, \pi^{2\tilde N}} \int dC d\lambda d\bpsi d\psi d\bvph d\vph \, e^{S},
\]
where $\tilde N=\sum_{i=1}^3 N_i$, $A=\int_{\mathbb{R}^{N_1N_2N_3}}  dC\, e^{-\alpha\, C^*\cdot C}$, and 
\s[
S=&-\alpha\, C^*\cdot C + \sum_{i=1}^3\left( \bpsi_{(i)}\cdot \psi_{(i)}+ \bvph_{(i)}\cdot \vph_{(i)}
+ I \lambda^*_{(i)} \cdot v_{(i)} +I \lambda_{(i)}\cdot v^*_{(i)} \right) \\
&-\sum_{(ijk)\in\{ (123)\}}\left( C^* \cdot \bpsi_{(i)} \vph_{(j)} v_{(k)}+C\cdot \bvph_{(i)} \psi_{(j)} v_{(k)}^*
+\frac{I}{2} C \cdot \lambda^*_{(i)}v_{(j)}^* v_{(k)}^*+\frac{I}{2} C^* \cdot \lambda_{(i)} v_{(j)} v_{(k)} \right) \\
&+\beta \sum_{i=2,3}\Big( 2 \bpsi_{(i)} \psi_{(i)}\cdot v_{(i)} v_{(i)} +2 \bvph_{(i)} \vph_{(i)} \cdot v_{(i)}^* v_{(i)}^* -
2 \bpsi_{(i)} \vph_{(i)} \cdot v_{(i)} v^*_{(i)} -2 \bvph_{(i)} \psi_{(i)} \cdot v_{(i)}^* v_{(i)}\\
&\hspace{2cm}+I (\lambda_{(i)}^* \cdot v_{(i)} -\lambda_{(i)} \cdot v_{(i)}^*)( v_{(i)}\cdot v_{(i)}- v_{(i)}^*\cdot v_{(i)}^*)
\Big).
\label{eq:bareaction}
\s]
Here $\sum_{(ijk)\in\{ (123)\}}$ is the summation over all the permutations of 123, namely,
 $(ijk)=$(123), (231), (312), (213), $\ldots$, and $\cdot$ indicates index contractions. For 
 instance,
 \s[
&C^*\cdot C=C^{*a_1 a_2 a_3} C_{a_1a_2a_3},\\
&\bpsi_{(i)}\cdot \psi_{(i)}= \bpsi_{(i)a} \psi_{(i)}^a,\\
&C^* \cdot \bpsi_{(2)} \vph_{(3)} v_{(1)}=C^{*a_1a_2a_3} \bpsi_{(2)a_2} \vph_{(3)a_3} v_{(1)a_1}, \\
&\bpsi_{(i)} \psi_{(i)}\cdot v_{(i)} v_{(i)} = \bpsi_{(i)a_1} \psi_{(i)}^{a_2} v_{(i)a_1} v_{(i) a_2}.
\label{eq:defcdot} 
\s]
Note that index contractions must be performed for the same index spaces, 
and also that some contractions coming from the gauge fixing terms are between upper (or lower)
indices, as on the last line.

The integration over $C$ and $\lambda$ is straightforward but lengthy. The details are explained in Appendix~\ref{app:comp}.
The result is
\[
\rhos(v)=\frac{\alpha^{\tilde N-2}\, e^{-\frac{\alpha}{|v|^2} }}{\beta^2 \pi^{\tilde N-2} |v|^{4(\tilde N-2)}} \delta^4(\cdots) Z_{\textrm{four-fermi}}
\label{eq:rhosfourfermi}
\]
where
\[
\delta^4(\cdots)=\prod_{i=2}^3 
\delta\left(|v_{(i)}|^2-|v_{(1)}|^2\right)
\delta\left( I(v_{(i)}\cdot v_{(i)}-v^*_{(i)}\cdot v^*_{(i)})\right),
\label{eq:deltafn}
\]
and 
\[
Z_{\textrm{four-fermi}}=\int d\bpsi d\psi d\bvph d\vph \, e^{S_{\textrm{four-fermi}}}
\label{eq:partfermi}
\]
with
\s[
S_{\textrm{four-fermi}}=&\sum_{i=1}^3 \left( \bpsi_{(i)}\cdot \psi_{(i)} +\bvph_{(i)} \cdot \vph_{(i)}\right) 
-\frac{1}{|v|^2}\sum_{i,j=1, i\neq j}^3
\left( v_{(i)}\cdot \bvph_{(i)}\, v_{(j)} \cdot \psi_{(j)}
+v_{(i)}^*\cdot \bpsi_{(i)}\, v_{(j)}^* \cdot  \vph_{(j)} \right)\\
&+g \sum_{i,j=1, i <  j}^3 \left(\bvph_{(i)}\psi_{(j)}+\bvph_{(j)}\psi_{(i)}\right) \cdot
I^{\perp}_{(i)}I^{\perp}_{(j)}\cdot \left(\bpsi_{(i)}\vph_{(j)}+\bpsi_{(j)}\vph_{(i)}\right) \\
& +
2 \beta \sum_{i=2,3} \left(  \bpsi_{(i)} \psi_{(i)}\cdot v_{(i)} v_{(i)}+ \bvph_{(i)} \vph_{(i)}\cdot v_{(i)}^* v_{(i)}^*
-\bpsi_{(i)} \vph_{(i)} \cdot v_{(i)} v^*_{(i)} -\bvph_{(i)} \psi_{(i)} \cdot v^*_{(i)} v_{(i)}
\right),
\label{eq:sfourfermi}
\s]
where 
\[
g=\frac{|v|^2}{\alpha},
\]
$I^\perp_{(i)}\ (i=1,2,3)$ denote the projectors to the transverse spaces against $v_{(i)}$ defined by
\[
I_{(i)a}^{\perp}{}^{a'} =\delta_{a}^{a'} -\frac{v_{(i)a } v^{*a'}_{(i)}}{|v|^2},
\label{eq:iperp}
\]
and $|v|=|v_{(i)}| \ (i=1,2,3)$ because of \eq{eq:deltafn}.
As can be seen in \eq{eq:sfourfermi}, the four-fermi interaction terms do not contain the components parallel to $v_{(i)}$,
similarly to the real cases \cite{Sasakura:2022zwc,Sasakura:2022iqd,Sasakura:2022axo}.

\section{Computation of the four-fermi theory}
\label{sec:compfermi}
In this section we explicitly compute the four-fermi partition function \eq{eq:partfermi} with \eq{eq:sfourfermi}.
As can be seen in \eq{eq:sfourfermi}, the fermion degrees of freedom can naturally be separated into
the components in the spaces spanned by $v_{(i)},v_{(i)}^*\ (i=1,2,3)$ 
and those transverse to them. Since $v_{(i)}$ and $v_{(i)}^*$
are generally linearly independent, the transverse spaces have dimensions $N_i-2$. So we take ${}^\forall N_i \geq 2$,
and separate the fermion degrees of freedom as
\s[
&\psi_{(i)}=\frac{v_{(i)}^*}{|v|} \psi_{(i)}^1
+\left( \frac{v_{(i)}}{|v|} -\frac{v_{(i)}^* v_{(i)}\cdot v_{(i)}}{|v|^3}\right) \psi_{(i)}^2+\psi_{(i)}^\perp,\\
&\bpsi_{(i)}=\frac{v_{(i)}}{|v|} \bpsi_{(i)}^1
+\frac{1}{F_{i}}\left( \frac{v^*_{(i)}}{|v|} -\frac{v_{(i)}v_{(i)}^*\cdot v_{(i)}^*}{|v|^3}\right) \bpsi_{(i)}^2+\bpsi_{(i)}^\perp,\\
&\vph_{(i)}=\frac{v_{(i)}}{|v|} \vph_{(i)}^1 
+\left( \frac{v^*_{(i)}}{|v|} -\frac{v_{(i)}v_{(i)}^*\cdot v_{(i)}^*}{|v|^3}\right) \vph_{(i)}^2+\vph_{(i)}^\perp, \\
&\bvph_{(i)}=\frac{v_{(i)}^*}{|v|} \bvph_{(i)}^1 
+\frac{1}{F_{i}}\left( \frac{v_{(i)}}{|v|} -\frac{v_{(i)}^* v_{(i)}\cdot v_{(i)}}{|v|^3}\right) \bvph_{(i)}^2+\bvph_{(i)}^\perp,\\
\label{eq:fermisep}
\s] 
where
\[
F_i=1-\frac{v_{(i)}\cdot v_{(i)} v^*_{(i)}\cdot v^*_{(i)}}{|v|^4}.
\]
Here, for a fermion $X$, $v_{(i)}\cdot X_{(i)}^\perp=v_{(i)}^* \cdot X_{(i)}^\perp=0$, and
$X_{(i)}^1$ and $X_{(i)}^2$ are one dimensional, while $X_{(i)}^\perp$ are 
$N-2$ dimensional. $X_{(i)}^1$ and $X_{(i)}^2$ should not be 
confused with the original $1,2$ components of $X_{(i)}$, 
but we will hereafter use this notation for simplicity, 
as there will be no confusions from contexts.

By the change of variables \eq{eq:fermisep} the kinetic term is transformed as
\[
\bar X_{(i)}\cdot X_{(i)}=\bar X_{(i)}^1 X_{(i)}^1+\bar X_{(i)}^2 X_{(i)}^2+\bar X_{(i)}^\perp\cdot X_{(i)}^\perp.
\] 
Therefore the fermion integration measure keeps the same form after the transformation of variables \eq{eq:fermisep}
(see \eq{eq:fermionintdef}).

By putting the decomposition \eq{eq:fermisep} into the four-fermi action \eq{eq:sfourfermi}, and after a 
straightforward computation, we obtain
\[
Z_{\textrm{four-fermi}}=\left. e^{\frac{\partial^2}{\partial k^2}} Z_{\rm quad}
\right |_{k_i^1=k_i^2=1, k_i^3=k_i^4=0}
\label{eq:zfermiquad}
\]
with
\[
Z_{\rm quad}=\int d\bpsi d\psi d\bvph d\vph \, e^{S_{\rm quad}},
\]
where
\s[
S_{\rm quad}=& \sum_{i=1}^3 \left( \bpsi_{(i)}^1 \psi_{(i)}^1+\bvph_{(i)}^1 \vph_{(i)}^1\right)
-\sum_{i,j=1,i\neq j}^3\left(\bvph_{(i)}^1 \psi_{(j)}^1+\bpsi_{(i)}^1 \vph_{(j)}^1 \right)\\
&+\sum_{i=1}^3 \Big( 
k_i^1 \left( \bpsi_{(i)}^2 \psi_{(i)}^2+\bpsi_{(i)}^\perp\cdot \psi_{(i)}^\perp  \right)
+k_i^2 \left( \bvph_{(i)}^2 \vph_{(i)}^2+\bvph_{(i)}^\perp \cdot \vph_{(i)}^\perp  \right)\\
&\hspace{2cm}+k_i^3  \left( \frac{1}{F_i} \bpsi_{(i)}^2 \bvph_{(i)}^2+\bpsi_{(i)}^\perp \cdot \bvph_{(i)}^\perp  \right)
+k_i^4 \left( F_i \,\psi_{(i)}^2 \vph_{(i)}^2 +\psi_{(i)}^\perp \cdot \vph_{(i)}^\perp  \right)
\Big) \\
&+2 \beta |v|^2 \sum_{i=2,3}
\left( \frac{v_{(i)}\cdot v_{(i)}}{|v|^2} \bpsi_{(i)}^1+\bpsi_{(i)}^2 
- \frac{v^*_{(i)}\cdot v^*_{(i)}}{|v|^2} \bvph_{(i)}^1-\bvph_{(i)}^2 \right) \left( \psi_{(i)}^1-\vph_{(i)}^1\right),
\label{eq:squad}
\s]
$k_{i}^{j}$ being auxiliary parameters, 
and 
\[
\frac{\partial^2}{\partial k^2} :=g  \sum_{i,j=1, i\neq j}^3 \left( 
\frac{\partial^2}{\partial k_i^3 \partial k_j^4}-\frac{\partial^2}{\partial k_i^1 \partial k_j^2}
\right).
\]
Here the four-fermi interaction terms  in \eq{eq:sfourfermi} have been rewritten by using the following identity
for fermions $\bar X,X, \bar Y,Y$,
\[
e^{g \frac{\partial^2}{\partial k \partial k'} } e^{k \bar X\cdot X +k' \bar Y\cdot Y}=
\sum_{n=0}^\infty \frac{g^n}{n!} (\bar X \cdot X)^n (\bar Y \cdot Y)^n e^{k \bar X\cdot X +k' \bar Y\cdot Y}
=e^{k \bar X \cdot X +k' \bar Y\cdot  Y+ g \bar X\cdot X  \bar Y\cdot  Y}.
\]
Note that the summation above is well-defined, because it is terminated at a finite order for fermions.

The 1,2 components and the $\perp$ components are decoupled from each other in \eq{eq:squad}, and therefore
can be computed 
separately. The fermion integrations are to compute Pfaffians from the quadratic terms. As for the 1,2 components,
we obtain from the explicit computation
\[
Z_{12}=-2^6 \beta^2 v_{(2)}\cdot  v_{(2)} v_{(3)}\cdot  v_{(3)} 
\prod_{i=1}^3 (k_i^1 k_i^2-k_i^3 k_i^4),
\label{eq:fermiint12}
\]
where we have used $v_{(i)}\cdot  v_{(i)}$ are real from \eq{eq:deltafn}.
As for the $\perp$ components, we obtain 
\[
Z_{\perp} = \prod_{i=1}^3 (k_i^1 k_i^2-k_i^3 k_i^4)^{N_i-2},
\]
where the power $N_i-2$ is because there are $N_i-2$ decoupled independent components.
Therefore, 
\[
Z_{\rm quad}=Z_{12}Z_{\perp}=-2^6 \beta^2 v_{(2)}\cdot  v_{(2)} v_{(3)}\cdot  v_{(3)} \prod_{i=1}^3 (k_i^1 k_i^2-k_i^3 k_i^4)^{N_i-1}.
\label{eq:zquad}
\]
By using the formula \eq{eq:qidentity} in Appendix~\ref{app:identity}, we obtain
\s[
&\left. e^{\frac{\partial^2}{\partial k^2}} \prod_{i=1}^3 (k_i^1 k_i^2-k_i^3 k_i^4)^{N_i-1}\right |_{k_i^1=k_i^2=1, k_i^3=k_i^4=0}
= \left( \prod_{i=1}^3 \Gamma(N_i) \right) F(g,\{N_i\}).
\label{eq:ekk}
\s]
Here  $F(g,\{N_i\})$ is a polynomial function defined by a generating function,
\[
F(g,\{ N_i \} )=\left. g^{\tilde N-3} (1-t_2+2 t_3)^{-2} \exp \left( \frac{t_1-2 t_2+3 t_3}{g(1-t_2+2 t_3)}\right) \right |_{\prod_{i=1}^3 l_i^{N_i-1}},
\label{eq:defofF}
\]
where
\s[
&t_1=l_1+l_2+l_3, \\
&t_2=l_1l_2+l_2l_3+l_3 l_1,\\
&t_3=l_1l_2 l_3,
\label{eq:ts}
\s]
and $|_{{l_i}^{N_i-1}}$ means taking the coefficient of the term of the order $\prod_{i=1}^3 {l_i}^{N_i-1}$ in the series 
expansion of \eq{eq:defofF} in $l_i$.
Therefore, collecting the results, \eq{eq:rhosfourfermi}, \eq{eq:zfermiquad}, \eq{eq:zquad}, and \eq{eq:ekk},
we obtain
\s[
\rhos(v)=&-\frac{2^6 \alpha^{\tilde N-2}e^{-\frac{\alpha}{|v|^2}} \prod_{i=1}^3 \Gamma(N_i)}{\pi^{\tilde N-2} |v|^{4(\tilde N-2)}} 
\, v_{(2)}\cdot  v_{(2)} v_{(3)}\cdot  v_{(3)}\, \delta^4(\cdots) \, F\left(\frac{|v|^2}{\alpha},\{N_i\}\right),
\label{eq:semifinal}
\s] 
where $\delta^4(\cdots)$ is given in \eq{eq:deltafn}.

\section{Gauge invariant distribution}
\label{sec:gaugeinv}
The signed distribution \eq{eq:semifinal} apparently depends on how we fix the gauge \eq{eq:gaugefix}. In this 
section we will extract the gauge invariant distribution from \eq{eq:semifinal}.

An obstacle for this purpose is that the Jacobian matrix $M(v,C,\beta)$ 
determining the quadratic terms of the fermions in \eq{eq:bareaction}
is dependent on the gauge fixing condition \eq{eq:gaugefix}. Therefore at first sight it does not seem possible to
obtain a gauge invariant distribution from \eq{eq:semifinal}. However, we can prove the following identity,
\[
{\rm Sign}\left( \det M(v,C,\beta) \right)= {\rm Sign}\left( v_{(2)}\cdot v_{(2)}\right) {\rm Sign} \left(v_{(3)}\cdot 
v_{(3)} \right) \, {\rm Sign} \left( {\rm det}_{\neq 0} M(v,C,\beta=0)\right),
\label{eq:signrelation}
\]
which is proven in Appendix~\ref{app:sign}.  Here $M(v,C,\beta=0)$ contains two zero eigenvalues coming from the 
two-dimensional continuous symmetry \eq{eq:rotation}, and the determinant on the righthand side is 
defined by the product of the non-zero eigenvalues of $M(v,C,\beta=0)$.
It is also straightforward to show that ${\rm det}_{\neq 0} M(v,C,\beta=0)$ is invariant under \eq{eq:rotation}.
Therefore, by simply multiplying 
${\rm Sign}\left( v_{(2)}\cdot v_{(2)}\right) {\rm Sign} \left(v_{(3)}\cdot v_{(3)} \right)$ to \eq{eq:semifinal}, 
the gauge dependence of the determinant can be removed.

The delta functions in \eq{eq:semifinal} still explicitly depend on the gauge fixing conditions. 
To remove this gauge dependence we smear \eq{eq:semifinal} along the symmetry orbit \eq{eq:rotation}.
Picking up the relevant part in \eq{eq:semifinal} and taking into account the multiplication of 
${\rm Sign} \left( v_{(i)}\cdot v_{(i)}\right)$ discussed above,  we obtain
\[
\frac{1}{2 \pi} \int_0^{2 \pi} d\theta_i \, {\rm Sign} \left( v_{(i)}\cdot v_{(i)}\right)  v_{(i)}\cdot v_{(i)}
\, \delta \left( I\left( v_{(i)}\cdot v_{(i)}\, e^{2 I \theta_i} -v_{(i)}^*\cdot v_{(i)}^*\, e^{-2 I \theta_i}\right)\right)=\frac{1}{2 \pi}.
\]
We must also correct the degeneracy 16 of the residual discrete gauge symmetry \eq{eq:residual}.
Thus, dividing \eq{eq:semifinal} by $16 (2 \pi)^2$, we obtain the gauge invariant distribution as 
\s[
\rho_{\rm signed}^{\rm gauge\, inv}(v)=&-\frac{\alpha^{\tilde N-2}e^{-\frac{\alpha}{|v|^2}} \prod_{i=1}^3 \Gamma(N_i)}{\pi^{\tilde N} |v|^{4(\tilde N-2)}} 
\delta\left(|v_{(2)}|^2-|v_{(1)}|^2)\delta(|v_{(3)}|^2-|v_{(1)}|^2\right) F\left(\frac{|v|^2}{\alpha},\{N_i\}\right).
\label{eq:rhogaugeinv}
\s] 

A comment is that this may not be the only gauge invariant distribution of the eigenvectors which are useful. 
The present definition counts the number of the eigenvectors by counting the number of the symmetry orbits 
\eq{eq:rotation}. This is obviously a natural way to count the number, but it seems also possible to 
define a quantity in which each orbit is counted with a weight such as its size. This weighted definition could be 
more natural, when one would be more interested in the volume eigenvectors fill, rather than the number. 
 
Since the distribution \eq{eq:rhogaugeinv} depends only on the common size $|v|=|v_{(i)}|$, it is 
more convenient to define the distribution of the square size $|v|^2$.   
We obtain 
\s[
\rho_{\rm signed}^{\rm square\, size} (|v|^2)
&= \int \prod_{i=1}^3 dv_{(i)} \delta\left( |v|^2-|v_{(1)}|^2 \right) \rho_{\rm signed}^{\rm gauge\, inv}(v) \\
&=-\frac{\alpha^{\tilde N-2}}{|v|^{2 (\tilde N-1)}}e^{-\frac{\alpha}{|v|^2}} F\left(\frac{|v|^2}{\alpha},\{ N_i \} \right).
\label{eq:distfinal}
\s]
The measure for \eq{eq:distfinal} as probabilities is $d|v|^2$.

\begin{figure}
\begin{center}
\includegraphics[width=10cm]{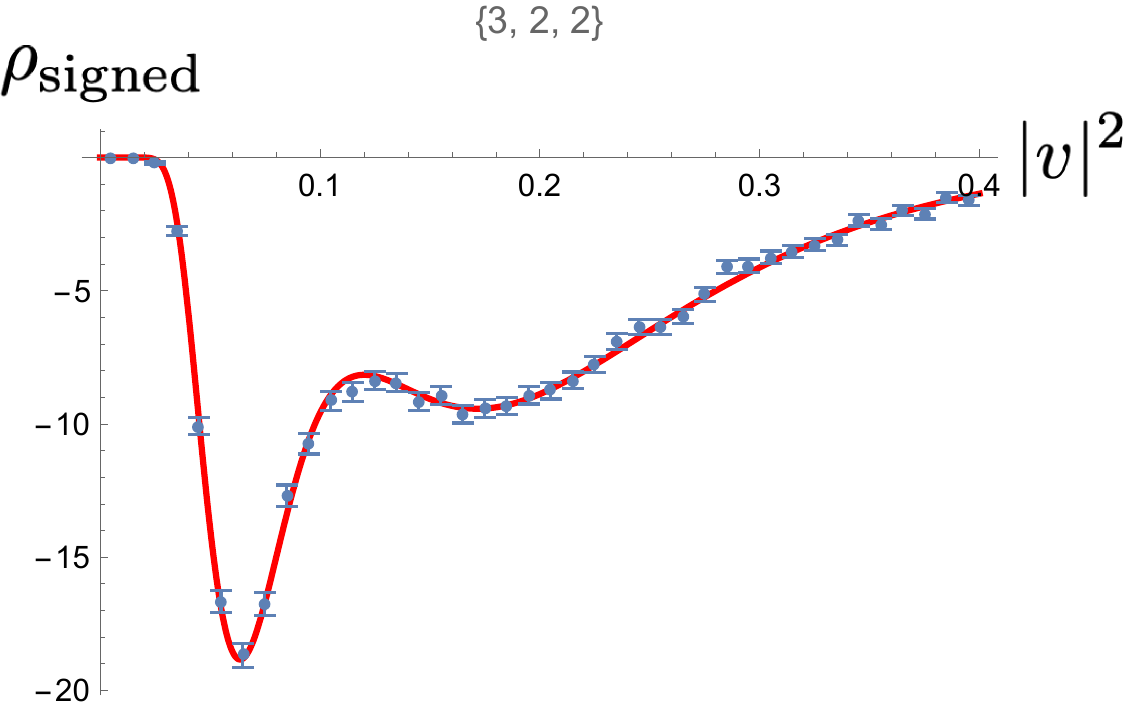}
\caption{A comparison between the Monte Carlo simulation (dots with error bars) 
and the analytic formula \eq{eq:distfinal} (solid line).
$\{ N_1,N_2,N_3\}=\{3,2,2\}$ and $N_{\rm samp}=10^4$. }
\label{fig:322}
\end{center}
\end{figure}

We crosschecked the final expression \eq{eq:distfinal} with Monte Carlo simulations. 
Although \eq{eq:distfinal} is gauge invariant, 
it is needed to employ a gauge fixing in the numerical simulations to uniquely solve the eigenvector equations \eq{eq:egeqs}. 
Rather than \eq{eq:gaugefix}, it is more convenient to take the gauge that $v_{(2)}^1,v_{(3)}^1$ (the first components 
of the vectors $v_{(2)},v_{(3)}$) be positive real.
The simulations were done by repeating randomly generating the real and imaginary parts of each component of $C$ 
by the normal distribution with mean value 0 and standard deviation 1 (this corresponds to setting $\alpha=1/2$), 
solving \eq{eq:egeqs} with the gauge fixing condition mentioned above,
and storing the size $|v_{(1)}|^2$ (it is always equal to $|v_{(2)}|^2$ and $|v_{(3)}|^2$)
and Sign(${\rm det}_{\neq 0} M(v,C,\beta=0)$).
Then we consider bins of size $\Delta$ for $|v|^2$, and processed the data to define
\s[
&{\cal N}_+\left(|v|^2\right)=\hbox{Number of solutions with 
square size in }[|v|^2-\Delta/2, |v|^2+\Delta/2)\hbox{ and positive sign},\\
&{\cal N}_-\left(|v|^2\right)=\hbox{Number of solutions with 
square size in }[|v|^2-\Delta/2, |v|^2+\Delta/2)\hbox{ and negative sign}.
\s]
Then the signed distribution from a simulation is computed by 
\s[
\rho^{\rm MC}_{\rm signed}\left(|v|^2\right)=\frac{1}{N_{\rm samp} \Delta}
\left( {\cal N}_+\left(|v|^2\right)-{\cal N}_-\left(|v|^2\right) \pm \sqrt{ {\cal N}_+\left(|v|^2\right)+{\cal N}_-\left(|v|^2\right) }
\right),
\s]
where $N_{\rm samp}$ is the total repetition number of random sampling of $C$ in a simulation, 
and the last term is an error estimate. 
An example for a comparison with the analytic formula \eq{eq:distfinal} is given in Figure~\ref{fig:322}. 
We performed similar analysis for $N_1\leq 4,N_2\leq3, N_3\leq2$ and $N_1=N_2=N_3=3$, 
and obtained good agreements supporting both the analytic formula and the numerical simulations.

We used a workstation which had a Xeon W2295 (3.0GHz, 18 cores), 128GB DDR4 memory, and Ubuntu 20 as OS.
The eigenvector equation \eq{eq:egeqs} was solved by the NSolve command of Mathematica 14. 

Lastly we point out a topological property of the signed distribution \eq{eq:distfinal}.
It is that the integration of \eq{eq:distfinal} over the whole region (namely, $\mathbb{R}_{\geq 0}$) is an integer.
Putting \eq{eq:defofF} into \eq{eq:distfinal} and performing the $|v|^2$ integration over $\mathbb{R}_{\geq 0}$, 
we obtain
\s[
\int_0^\infty d|v|^2\, \rho_\text{signed}^\text{square size}(|v|^2)&=-\left. \int_0^\infty dx\, (1-t_2+2 t_3)^{-2} \exp \left(- x \frac{1-t_1+ t_2- t_3}{1-t_2+2 t_3}\right)\right |_{\prod_{i=1}^3 l_i^{N_i-1}} \\
&=-\left. \frac{1}{(1-t_1+ t_2- t_3)(1-t_2+2 t_3)}\right |_{\prod_{i=1}^3 l_i^{N_i-1}}.
\s]
Since $(1-x)^{-1}=1+x+x^2+\cdots$, the series expansion in $l_i$ has integer coefficients\footnote{One may suspect that
the number may be all negative. This is not true. In fact, for instance, it is positive for $(N_1,N_2,N_3)=(4,4,4)$.}. 

\section{Large-\!$N$ limit}
\label{sec:largen}
In this section we take the large-\!$N$ limit of our results\footnote{Here $N$ represents 
the order of $N_i$. Namely, $N$ is a large number with $N_i/N\sim {\cal O}(1)$.}.  
The overall properties turn out to be very similar to the previous case of the real symmetric random 
tensor \cite{Kloos:2024hvy}.
We find a critical point, denoted by $|v|_c$, and also the end of the distribution, denoted 
by $|v|_{\rm end}\ (<|v|_c)$. 
The critical point $|v|_c$ separates the two regions where the signed distribution behaves differently: At $|v|<|v|_c$
it is monotonous, while it is infinitely oscillatory at $|v|>|v|_c$ in the large-\!$N$ limit. 

We perform the analysis in two independent ways. 
In Section~\ref{sec:sd}, we perform an analysis of the four-fermi theory \eq{eq:sfourfermi} by using 
the Schwinger-Dyson (SD) equation  
in the leading order of $N$. For $N_2=N_3$, we derive a cubic equation determining $|v|_c$.
The critical point $|v|_c$ separates the following two regions: 
For $|v|<|v|_c$ the solution to the SD equation is real, and so is the effective action, while for $|v|>|v|_c$ they are complex. 
The former leads to a monotonous behavior, while the latter suggests oscillatory behavior. 
We also derive the equation determining the end of the signed distribution, denoted by $|v|_{\rm end}$,
and numerically compute the values for some cases.   
In Section~\ref{sec:largenfinal}, we perform a saddle point analysis of an integral representation of 
\eq{eq:distfinal} for $N_1=N_2=N_3$.
We obtain the same results of $|v|_c,|v|_{\rm end}$ as those by the SD method.
For $|v|<|v|_c$ there is only one relevant saddle on the real axis, while for $|v|>|v|_c$ there are two relevant complex 
saddles conjugate to each other. These confirm the behavior suggested by the SD method. 

\subsection{Schwinger-Dyson method}
\label{sec:sd}
In this subsection we analyze the four-fermi theory \eq{eq:sfourfermi} by a method using 
the Schwinger-Dyson equation in the leading order of $N$. The method was previously used for 
the case of the real symmetric random tensor \cite{Sasakura:2022iqd,Kloos:2024hvy}.  
For this paper to be self-contained, the method is explained in some detail in Appendix~\ref{app:methodsd}.
 
In the large-\!$N$ limit the transverse components in \eq{eq:sfourfermi} will dominate the dynamics, 
so we ignore the field components parallel to $v_{(i)},v_{(i)}^*$ for simplicity.
The transverse part of \eq{eq:sfourfermi} is given by
\s[
S_\perp &=\sum_{i=1}^3 \left( \bpsi_{(i)}\cdot \psi_{(i)} +\bvph_{(i)}\cdot \vph_{(i)}\right) 
+g \sum_{i,j=1, i> j}^3 \left(\bvph_{(i)}\psi_{(j)}+\bvph_{(j)}\psi_{(i)}\right) \cdot
 \left(\bpsi_{(i)}\vph_{(j)}+\bpsi_{(j)}\vph_{(i)}\right) \\
 &= \sum_{i=1}^3 \left( \bpsi_{(i)}\cdot \psi_{(i)} +\bvph_{(i)} \cdot \vph_{(i)}\right) 
 -g \sum_{i,j=1,i\neq j}^3 
 \left( 
 \bar\psi_{(i)}\cdot \psi_{(i)} \, \bar \varphi_{(j)} \cdot  \varphi_{(j)}
 + \bar\psi_{(i)}\cdot \bar \varphi_{(i)} \, \varphi_{(j)}\cdot \psi_{(j)}
 \right).
 \label{eq:strans}
\s]
Here the fields have only transverse components and therefore are $N_i-2$ dimensional, 
but we simply regard them as $N_i$ for notational simplicity in the following discussions, 
because we are considering the leading order of $N$.

The starting point of the SD method is to assume two-fermion expectation values as
\s[
&\langle \bar \psi_{(i)a} \psi_{(i)}^b\rangle  =Q^i_1 \delta_{a}^b,  \\
&\langle \bar \psi_{(i)a} \bar \varphi_{(i)}^b\rangle =Q^i_2 \delta_{a}^b, \\
&\langle  \varphi_{(i)a} \psi_{(i)}^b \rangle=Q^i_3 \delta_{a}^b, \\
&\langle \varphi_{(i)a} \bar\varphi_{(i)}^b \rangle=Q^i_4 \delta_{a}^b,  \\
&\hbox{the others}=0,
\label{eq:qassume}
\s]
where $\langle {\cal O}\rangle =\int d\bar \psi d\psi d\bar \varphi d\varphi\, {\cal O}\, e^{S_\perp}/
\int d\bar \psi d\psi d\bar \varphi d\varphi\, e^{S_\perp}$, and 
$Q^i_j$ are some values to be determined.  The form of \eq{eq:qassume} is the most general one
satisfying the symmetry possessed by  
\eq{eq:strans}\footnote{The fermion integration measure is also invariant under the symmetry.}, 
\[
\bar \psi_{(i)a}'= \bar \psi_{(i)a'}G_{(i)}^{a'}{}_a ,\  \varphi_{(i)a}'=  \varphi_{(i)a'}G_{(i)}^{a'}{}_a ,\ 
\psi'{}_{(i)}^{a}=G^{-1}_{(i)}{}^{a}{}_{a'}  \psi_{(i)}^{a'},\  \bar\varphi'{}_{(i)}^{a}=G^{-1}_{(i)}{}^{a}{}_{a'}  \bar\varphi_{(i)}^{a'},
\]
where $G_{(i)} \in GL(N_i)$. Then, assuming $g\sim 1/N$ and following the procedure in Appendix~\ref{app:methodsd},  
we obtain an effective action in the leading order as 
\s[
S_{\rm eff}&=\langle S_{\rm \perp} \rangle -\sum_{i=1}^3 N_i \log (-\det Q_{(i)}) \\
&= \sum_{i=1}^3 N_i (Q^i_1-Q^i_4)  +g \sum_{i,j=1, i\neq j}^3 N_i N_j (Q^i_1 Q^j_4-Q_2^i Q_3^j)
 -\sum_{i=1}^3 N_i \log\left(- \det Q_{(i)}\right),
\s]
where $Q_{(i)}$ are the 2 by 2 matrices defined by 
\[
Q_{(i)}=\left(
\begin{array}{cc}
Q_1^i & Q_2^i \\
Q_3^i & Q_4^i
\end{array}
\right).
\]
Here the computation of $\langle S_{\rm \perp} \rangle$ using the assumption \eq{eq:qassume} 
has been performed in the leading order. For instance,  
\s[
\langle \bar\psi_{(i)}\cdot \psi_{(i)} \, \bar \varphi_{(j)} \cdot  \varphi_{(j)}\rangle &\sim 
\langle \bar\psi_{(i)}\cdot \psi_{(i)}\rangle \langle \bar \varphi_{(j)} \cdot  \varphi_{(j)}\rangle 
=N_i Q_1^i N_j Q_4^j.
\s]

As shown in Appendix~\ref{app:stationary}, in the weak coupling region of $g$,  
the stationary conditions $\partial S_{\rm eff}/\partial Q^i_j=0$ reduce to finding a real solution $q$
to the equation, 
\s[
1+2q-\sum_{i=1}^3 \sqrt{1+4 g N_i q^2}=0.
\label{eq:q}
\s]
With the solution $q$, $Q^i_j$ are expressed as 
\[
Q^i_1=-Q_4^i=\frac{1}{2 q g N_i}\left(\sqrt{1+4 g N_i q^2} -1\right),\  Q^i_2=Q_3^i=0.
\label{eq:qsol}
\]

When $g$ is small enough, \eq{eq:q} has a real solution. The solution continues to exist until $g$ reaches 
a critical value $g_c$. At $g>g_c$ there are no real solutions. The critical value depends on $N_i$. 
For $N_i$ being all equal, namely, $N_i=N(=\tilde N/3)$, one can easily obtain
$g_c=1/(8 N)$, or
\[
|v|_c=\sqrt{\frac{ \alpha}{8 N} }\hbox{ for } N_i=N.
\label{eq:criticalv2}
\]

It seems difficult to obtain $g_c$ for general $N_i$. But for $N_2=N_3$, we obtain  
a cubic equation which determines $g_c$ from the condition for $q$ to be real:
\[
1 + 6 g_c(N_1 - 2 N_2)  - 
 3 g_c^2 (5 N_1^2 + 16 N_1 N_2 - 16 N_2^2)+ 8 g_c^3 (N_1 - 2 N_2)^3=0.
 \label{eq:gcn2}
\]

For $g>g_c$, the solution becomes complex and so is the value of $S_{\rm eff}$. One would probably have to
sum up the contributions of two complex solutions to obtain a real distribution, 
but it seems rather ambiguous how it should be done in the present SD framework. 
In Section~\ref{sec:largenfinal}, we will perform a saddle point analysis of \eq{eq:distfinal}, 
which is more definite.

The end of the distribution can also be obtained by the SD method.  We need to take into account the multiplicative factors 
in \eq{eq:rhosfourfermi} as well as the contribution of the volumes of $v_{(i)}$.  
As for the latter, the details of the delta functions in \eq{eq:rhosfourfermi} can be ignored in the leading order,  
and we just have to multiply the surface volumes $2 \pi^{N_i} |v|^{2 N_i-1}/\Gamma(N_i)$ of the $2N_i-1$-dimensional 
spheres of radiuses $|v_i|=|v|$. 
By introducing new parameters, 
$n_i=N_i/\tilde N,\ \tilde g=g \tilde N$, we obtain $\rho_{\rm signed}^{\rm square\, size} 
\sim e^{\tilde N \tilde S}$, where 
\[
\tilde S=
-\log \tilde g -\frac{1}{\tilde g} +1 -\sum_{i=1}^3 n_i \log n_i+\tilde S_{\rm eff} -\tilde S_{\rm eff}(Q_i^j(g=0), g=0) 
\]  
with
\[
\tilde S_{\rm eff}= \sum_{i=1}^3 n_i (Q^i_1-Q^i_4)  +\tilde g \sum_{i,j=1, i\neq j}^3 n_i n_j (Q^i_1 Q^j_4-Q_2^i Q_3^j)
 -\sum_{i=1}^3 n_i \log(- \det Q_{(i)}),
\]
where $Q_i^j$ are the solutions in \eq{eq:qsol}.
Here the subtraction of $\tilde S_{\rm eff}(Q_i^j(g=0), g=0)=2$ 
is because the fermion partition function is normalized\footnote{This normalization is
included in the formula $\det M=\int d\bar \psi d\psi \, e^{\bar \psi M \psi}$ \cite{zinn}.} 
as $Z_\text{four-fermi}(g=0)=1$ (See \eq{eq:appzeff}). 
Then the end of the distribution is given by the solution to $\tilde S=0$. 
Because of the logarithmic term, the solution can only be obtained numerically. 
For $N_i=N(=\tilde N/3)$, we obtain $\tilde g_{\rm end}\sim 0.364213$,
or
\[
|v|_{\rm end}\sim 0.348431\sqrt{ \frac{\alpha}{N} } \hbox{ for } N_i=N.
\label{eq:vend}
\]

\begin{figure}
\begin{center}
\includegraphics[width=7cm]{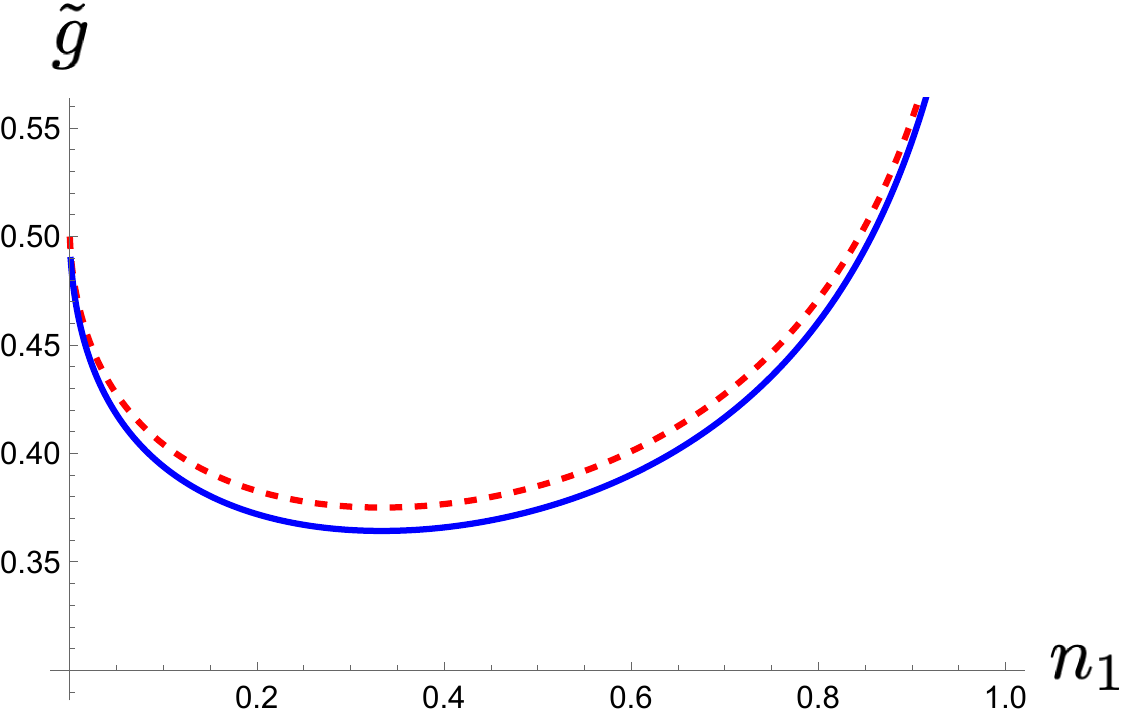}
\caption{The values of $\tilde g_c$ (broken line) and $\tilde g_{\rm end}$ (solid line) 
for $N_1=n_1 \tilde N,N_2=N_3=(1-n_1)\tilde N/2$.
They take $1/2$ and $1$ at $n_1=0$ and $n_1=1$, respectively.}
\label{fig:gt}
\end{center}
\end{figure}
It would be interesting to see the behavior of $\tilde g_c$ and  $\tilde g_{\rm end}$ for $N_2=N_3$
by \eq{eq:gcn2} and  numerically solving the condition $\tilde S=0$, respectively.
This is drawn in Figure~\ref{fig:gt}.

\subsection{Large-\!$N$ limit of \eq{eq:distfinal}}
\label{sec:largenfinal}

In this subsection we take the large-\!$N$ limit of the expression \eq{eq:distfinal}. For simplicity we consider only the 
case with equal $N_i$, namely, $N_i=\tilde N/3$.

We may replace $|_{\prod_{i=1}^3 l_i^{N_i-1}}$ in \eq{eq:defofF} by contour integrals around the origin,
\[
|_{\prod_{i=1}^3 l_i^{N_i-1}} \rightarrow \frac{1}{(2 \pi I)^3} \prod_{i=1}^3 \oint dl_i \frac{1}{l_i{}^{N_i}}.  
\]
We further assume $|v|\sim 1/\sqrt{N}$. Then we can rewrite the distribution \eq{eq:distfinal} as 
\[
\rho_{\rm signed}^{\rm square\ size} (|v|^2)=-\tilde N^2 \alpha^{-1} x^2  \frac{1}{(2 \pi I)^3} 
\prod_{i=1}^3 \oint dl_i \, (1-t_2+2 t_3)^{-2} e^{ \tilde N h}, 
\label{eq:rhointbyl}
\]
where $x=\alpha/(\tilde N |v|^2)$, and 
\[
h=-\frac{1}{3} \sum_{i=1}^3 \log l_i -x \frac{1-t_1+t_2-t_3}{1-t_2+2 t_3}.
\]
Here the factor $1/3$ is because of $N_i/\tilde N=1/3$ for the present case. 
Therefore in the large-\!$N$ limit the distribution around $|v|\sim 1/\sqrt{N}$ 
can well be approximated by the saddle point analysis of $h$. 

The analysis of saddle points of $h$ is elementary and we obtain the following saddle points:
\s[
&(1) \ l_1=l_2=l_3=\frac{1}{8} \left(-4 + 3 x - \sqrt{3} \sqrt{x (-8 + 3 x)}\right),\\
&(2)\ l_1=l_2=l_3=\frac{1}{8} \left(-4 + 3 x + \sqrt{3} \sqrt{x (-8 + 3 x)}\right),\\
&(3)\ l_1=\frac{1}{3x},\ l_2=l_3=3 x, \hbox{ and exchanged ones}.
\label{eq:criticalpoints}
\s]
There is a critical point at $x_c=8/3$, which indeed agrees with \eq{eq:criticalv2}.

Lefschetz thimbles are the steepest descent contours passing through saddle points\footnote{
See for example Section 3 of \cite{Witten:2010cx} for 
the steepest descent method in terms of Lefschetz thimbles.}.
We can determine the relevant saddle points out of \eq{eq:criticalpoints} 
by finding how the original integration contours in \eq{eq:rhointbyl} can be deformed as a sum of Lefschetz thimbles. 
In the present case, this criterion can be checked rather clearly for $x>x_c$.
For $x>x_c$ the saddle points are on the real axis, and the eigenvalues of the Hessian matrices
on these saddle points are real. A contour around the origin can naturally be deformed to 
a path crossing the real axis on a saddle point, if the Hessian eigenvalue is positive there, 
because the path then becomes a descent one.   
From an explicit computation of the Hessian eigenvalues, this criterion is satisfied for all $l_i$
only for the saddle point (1) above. Therefore for $x>x_c$ the relevant saddle point is (1), and the others are not.

As for the region $x<x_c$, the Hessian matrices take complex values for (1) and (2), so we need more careful analysis.
This can be done by following the saddle point (1) as $x$ is decreased from large values.
At $x=x_c$ the two saddle points (1) and (2)
meet and become complex at $x<x_c$. Therefore they both satisfy the criterion at least in the vicinity of $x=x_c$, and would
be expected to continue so for $x<x_c$\footnote{However, when $x$ substantially departs from $x_c$, there may occur
Stokes phenomenon, which rearranges the relevant saddles. Here we restrict ourselves to $x$ being still close to $x_c$.}. 
The saddle points (3) do not satisfy the criterion for all the values of $x$.  

The correctness of the choice of the relevant saddle points above can be checked by doing the saddle point approximation
of \eq{eq:rhointbyl}.
Taking into account the quadratic order deviations around the saddle points leads to 
\[
\rho_{\rm signed}^{\rm square\, size}(|v|^2)\sim -(2 \pi )^{-3/2} \tilde N^{1/2}  x^2 
\sum_{i=1\ { \rm for } \ x>x_c \atop i=1,2\ { \rm for } \ x<x_c}
\left(1-3 \left(l^{(i)}\right)^2+2 \left(l^{(i)}\right)^3\right)^{-2}  \left( \det h_2^{(i)} \right)^{-1/2} e^{\tilde N h^{(i)}_0},
\]
where $l^{(i)}(=l_1=l_2=l_3)$ is the location of the saddle point $(i)$ in \eq{eq:criticalpoints}, 
and $h_0^{(i)}$ and $h_2^{(i)}$ are the values of 
$h$ and the Hessian matrix there, respectively. The explicit forms of these are given in Appendix~\ref{app:h0h2}.
As shown in Figure~\ref{fig:app}, we obtain fairly good agreement between the saddle point approximation 
and the exact expression.

\begin{figure}
\begin{center}
\includegraphics[width=7cm]{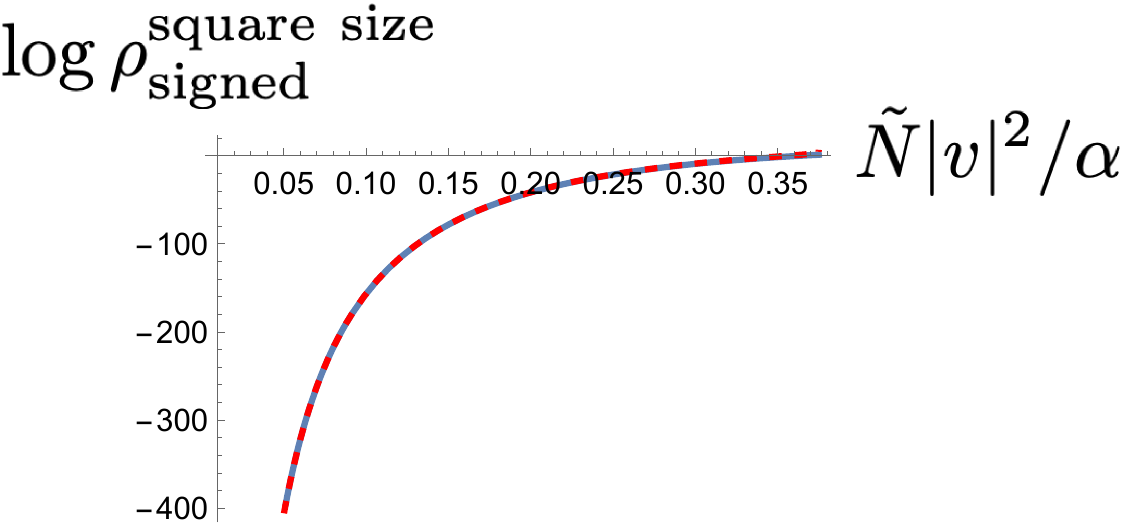}
\hfil
\includegraphics[width=7cm]{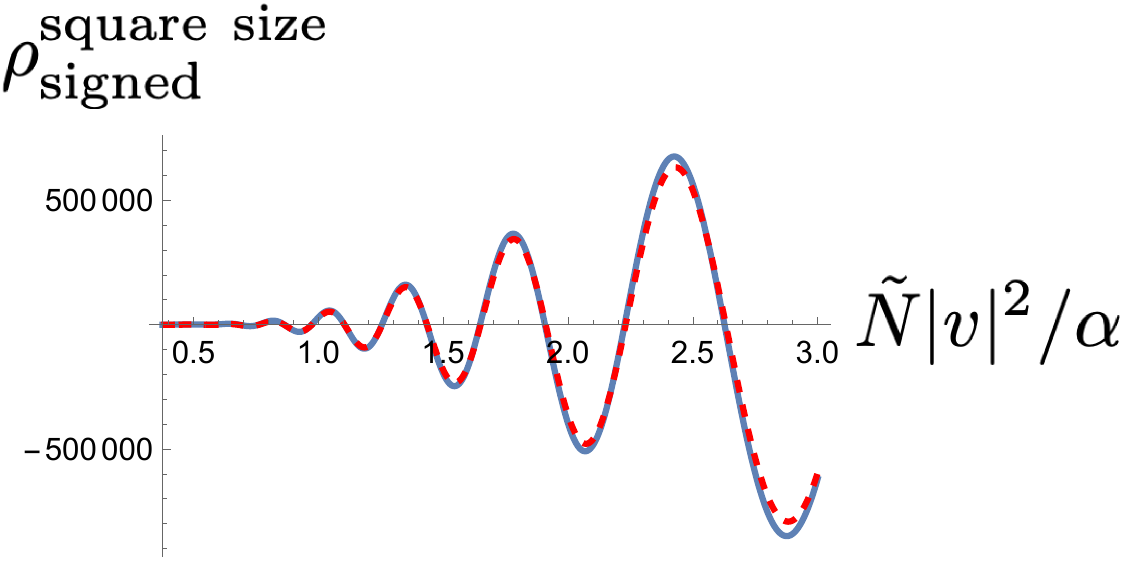}
\caption{The exact (solid line) and the large-\!$N$ (broken line) expressions are compared in the two regions $x>x_c$ (left)
and $x<x_c$ (right) for $N=9$ and $\alpha=1/2$. 
They are in good agreement.}
\label{fig:app}
\end{center}
\end{figure}

The end of the distribution can also be computed by the present treatment. Since the overall factor
of \eq{eq:rhointbyl} does not contain any exponential behavior in $\tilde N$, the condition is given by
\[
h=0
\label{eq:hzero}
\]
in the region $x>x_c$, where the saddle point value (1) should be inserted for $l_i$ . We find $x_\text{end}=2.74564$, 
which indeed leads to the same result \eq{eq:vend}.

\section{Some applications}
\label{sec:app}
In this section we explain a few applications of $|v|_{\rm end}$ obtained in Section~\ref{sec:largen}. 
For the applications, we need to make the following two assumptions, which are known to be true for the case of 
the real symmetric random tensor \cite{parisi,example,randommat,secondmoment}.

\begin{figure}
\begin{center}
\includegraphics[width=7cm]{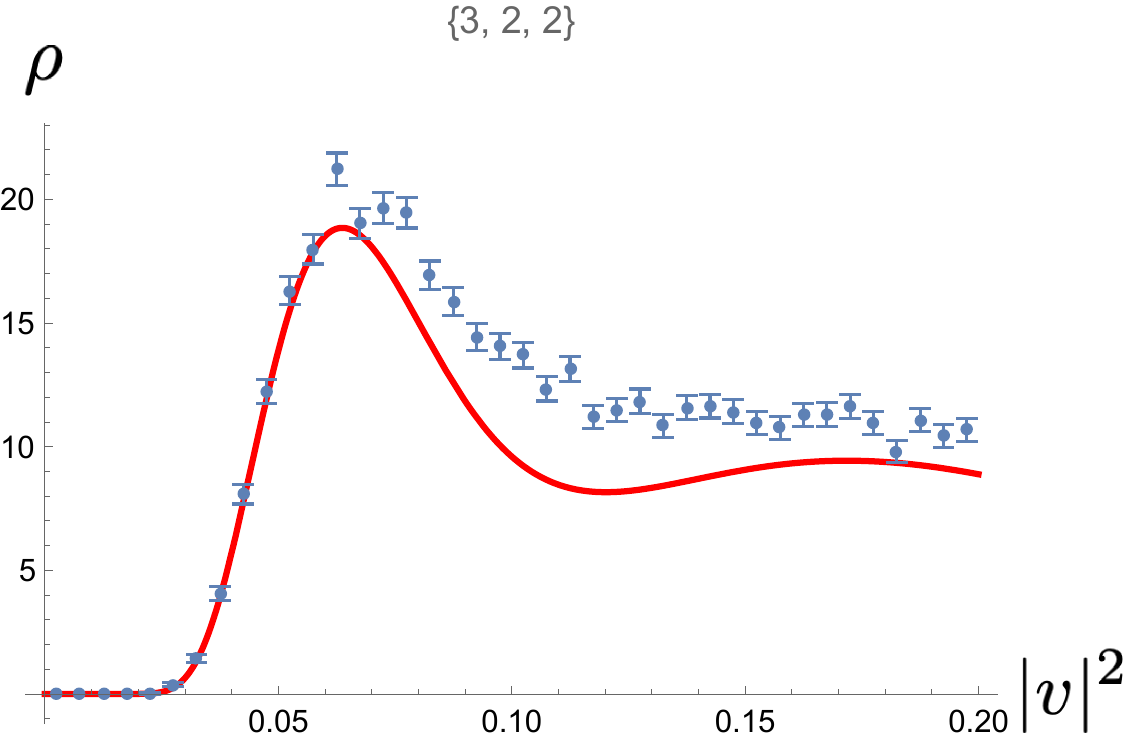}
\caption{A comparison between the (minus of) signed distribution $-\rho_{\rm signed}^{\rm square\, size}$ (solid line)
and the genuine distribution $\rho$ (dots with an error bar) by MC with $N_{\rm rep}=10^4$ for $(N_1,N_2,N_3)=(3,2,2)$.
They agree well in the vicinity of  the ends of the distributions.}
\label{fig:compare}
\end{center}
\end{figure}

The first one is that the singed and the genuine distributions have a common end in the large-\!$N$ limit.
As discussed in Section~\ref{sec:strategy}, this is naturally expected, 
because the set of the eigenvector equations \eq{eq:egeqs} can be derived as the equation of stationary points 
of a potential. One can also explicitly see the coincidence
by Monte Carlo simulations as shown in an example in Figure~\ref{fig:compare}.

The other assumption is that the mean of the eigenvector distribution defined in \eq{eq:genuine} 
can be identified with the distribution itself in the large-\!$N$ limit.
In other words, the fluctuation of the distribution vanishes in the large-\!$N$ limit. 
In fact this has been proven for the real symmetric random tensor in \cite{secondmoment}.
To our knowledge, a similar statement has not been proven for the complex random tensor.

Under the two assumptions above, we can identify $|v|_{\rm end}$ in \eq{eq:vend}, 
which has been computed from the signed 
distribution, as the end of the genuine distribution in the large-\!$N$ limit. 
We assume this for the applications below. 

\subsection{Largest eigenvalue}
The eigenvalue $\lambda$ of a complex order-three tensor \cite{qibook}  can be defined by normalizing the eigenvectors 
in \eq{eq:egeqs}:
\s[
&C_{a_1a_2a_3} w_{(2)}^{*a_2} w_{(3)}^{*a_3}=\lambda w_{(1) a_1},\\
&C_{a_1a_2a_3} w_{(3)}^{*a_3} w_{(1)}^{*a_1}=\lambda w_{(2) a_2},\\
&C_{a_1a_2a_3} w_{(1)}^{*a_1} w_{(2)}^{*a_2}=\lambda w_{(3) a_3},
\label{eq:eigen}
\s] 
where $|w_{(1)}|^2=|w_{(2)}|^2=|w_{(3)}|^2=1$ and $\lambda=1/|v|$.
Therefore  the largest eigenvalue in the large-\!$N$ limit is given by
\[
\lambda_{\rm max}= \frac{1}{|v|_{\rm end}}.
\label{eq:lammaxgen}
\]
In particular, from \eq{eq:vend}, we obtain
\[
\lambda_{\rm max} \sim 2.87001 \sqrt{\frac{N}{\alpha}} \hbox{ for } {N_i=N}.
\label{eq:lammax}
\]

\subsection{Geometric measure of entanglement}
In quantum information theory the amount of entanglement of a state is an interesting quantity. 
Let us consider an arbitrary state $|\Psi\rangle\in {\cal H}_1 \otimes {\cal H}_2\otimes {\cal H}_3$ in the Hilbert space of a
product of three Hilbert spaces. The amount of entanglement of $|\Psi \rangle$ 
would be characterized by the geometric measure of entanglement defined by the smallest distance from 
the separable states \cite{shi,barnum,estimate},
\[
D_{\rm E}(|\Psi\rangle)={\rm Min}_{|\psi_1\rangle, |\psi_2\rangle, |\psi_3\rangle}
\big| |\Psi\rangle-|\psi_1\rangle\otimes |\psi_2\rangle \otimes |\psi_3\rangle \big |,
\label{eq:de}
\]
where $\big| |X \rangle \big|=\sqrt{\langle X | X \rangle}$, and $\big| |\Psi \rangle \big|=\big| |\psi_i \rangle \big|=1$.
An interesting question is the typical value of $D_{\rm E}$ when $|\Psi\rangle$ is given randomly.

To relate this quantity to our computation, let us consider the square of the righthand side of \eq{eq:de}. We obtain 
\[
\big| |\Psi\rangle-|\psi_1\rangle\otimes |\psi_2\rangle \otimes |\psi_3\rangle \big |^2=2-\sigma-\sigma^*,
\label{eq:distsq}
\]
where
\[
\sigma=\langle \Psi |\, |\psi_1\rangle\otimes |\psi_2\rangle \otimes |\psi_3\rangle.
\]
Obtaining the smallest value of \eq{eq:distsq} is to find the largest value of ${\rm Re} [\sigma]$.
By expressing these states as $|\Psi\rangle=C_{a_1a_2a_3} | a_1\rangle \otimes | a_2\rangle \otimes | a_3 \rangle$
and $|\psi_i\rangle=w_{(i)a_i} |a_i\rangle$, we obtain $\sigma=C^{*a_1a_2a_3} w_{a_1}w_{a_2}w_{a_3}$.
The stationary condition of $\sigma$ under the normalization constraints $|w_{(i)}|^2=1$\
gives the eigenvalue equations \eq{eq:eigen}\footnote{One may use the method of the Lagrange multiplier to
derive it.}. The largest eigenvalue $\lambda_{\rm max}$ gives the largest $\sigma$.

In our system, the tensor $C$ is not normalized but we can normalize the size in average. 
The average of the size is $\sqrt{\langle |C|^2 \rangle}=\sqrt{N_1 N_2 N_3/\alpha}$ for the Gaussian distribution 
$e^{-\alpha C^*C}$. 
In particular, for $N_i=N$, we obtain
\[
\sigma_{\rm max} \sim \frac{\lambda_{\rm max}}{\sqrt{\langle |C|^2 \rangle}} \sim 
\frac{2.87001}{N} \hbox{ for }N_i=N,
\label{eq:smax}
\]  
from \eq{eq:lammax}.

Reference \cite{estimate} 
studied the geometric measure of entanglement by performing numerical simulations. Their value $C_0$
corresponds to $\lambda_{\rm max}$ for $\alpha=N/2$. From \eq{eq:lammax} we obtain
$\lambda_{\rm max}\sim 4.0588$ for this case, and the value seems consistent with their numerically computed 
value $C_0\sim 4.143529$ with a $2 \%$ difference, which is smaller than the error they report for the real symmetric case.    

\subsection{The best rank-one approximation}
The best rank-one approximation \cite{SAPM:SAPM192761164,Carroll1970,bestrankone, comon} 
of a complex order-three tensor is to consider the following minimization problem,
\[
{\rm Min}_{\phi} (C^{*a_1a_2a_3}-\phi_{(1)}^{*a_1} \phi_{(2)}^{*a_2} \phi_{(3)}^{*a_3})
(C_{a_1a_2a_3}-\phi_{(1)a_1} \phi_{(2)a_2} \phi_{(3)a_3}),
\]
where $\phi_{(i)}$ are the vectors in the $i$-th index space. The stationary condition gives
\s[
&C_{a_1a_2a_2} \phi^{*a_2}_{(2)} \phi^{*a_3}_{(3)}=|\phi_{(2)}|^2 |\phi_{(3)}|^2  \phi_{(1)a_1},\\
&C_{a_1a_2a_2} \phi^{*a_3}_{(3)} \phi^{*a_1}_{(1)}=|\phi_{(3)}|^2 |\phi_{(1)}|^2  \phi_{(2)a_2},\\
&C_{a_1a_2a_2} \phi^{*a_1}_{(1)} \phi^{*a_2}_{(2)}=|\phi_{(1)}|^2 |\phi_{(2)}|^2  \phi_{(3)a_3}.
\label{eq:eqphi}
\s]

It is obvious that one can take a gauge $|\phi_{(1)}|=|\phi_{(2)}|=|\phi_{(3)}| (=|\phi|)$, since only the product 
$\phi_{(1)} \phi_{(2)} \phi_{(3)}$ matters. By taking the gauge and introducing $w_{(i)}=\phi_{(i)}/|\phi|$ 
with appropriate phases, \eq{eq:eqphi} can be rewritten into the eigenvalue equations \eq{eq:eigen} with $\lambda=|\phi|^3$.
For such a solution 
\[
(C^{*a_1a_2a_3}-\phi_{(1)}^{*a_1} \phi_{(2)}^{*a_2} \phi_{(3)}^{*a_3})
(C_{a_1a_2a_3}-\phi_{(1)a_1} \phi_{(2)a_2} \phi_{(3)a_3})= |C|^2-|\phi|^6=|C|^2-\lambda^2.
\] 
Therefore, the best rank-one approximation is given by the eigenvector with the largest eigenvalue 
$\lambda_{\rm max}$.
The corresponding rank-one tensor $\phi_{(1)}^{\rm max} \otimes \phi_{(2)}^{\rm max}\otimes \phi_{(3)}^{\rm max}$ 
has size $|\phi^{\rm max}|^3=\lambda_{\rm max}$.
Therefore the relative size of the rank-one tensor to the size of $C$ in the 
the best rank-one approximation is given by
\[
\frac{|\phi_{(1)}^{\rm max} \otimes \phi_{(2)}^{\rm max}\otimes \phi_{(3)}^{\rm max}|} {\sqrt{\langle |C|^2 \rangle}}\sim 
\frac{2.87001}{N} \hbox{ for }N_i=N.
\]

\section{Summary and future prospects}
\label{sec:summary}
In this paper we have computed the signed eigenvalue/vector distribution of the complex order-three random tensor
by computing a partition function of a four-fermi theory. The final result
has a compact expression in terms of a generating function, which has an expansion
whose powers are the dimensions of the index spaces.
We have studied the large-\!$N$ limit both by the Schwinger-Dyson equation and by a saddle point analysis,
and have obtained the critical point where the behavior of the signed distribution qualitatively changes 
and the end of the signed distribution in the limit.
Under plausible assumptions, we applied the location of the end to compute
the largest eigenvalue, the geometric measure of entanglement, and 
the best rank-one approximation in the large-\!$N$ limit. 

A new challenge compared to the previous cases was the presence of a continuous degeneracy of the solutions 
to the eigenvalue/vector equations, which is generated by a two-dimensional Abelian symmetry. 
To apply the procedure to rewrite the distribution as a partition function of a quantum field theory, 
we modified the eigenvalue/vector equations so that they included the gauge fixing condition to remove the degeneracy, 
and have extracted the gauge invariant distribution at the final stage.
This procedure we took, however, appears to be specific to this particular case. 
It would be better to have a more systematic general procedure 
as in the gauge fixing procedure in gauge theories.

The applications in Section~\ref{sec:app} are based on the following two assumptions.
One is that the signed and the genuine distributions have a common end in the large-\!$N$ limit. 
The other is that the distribution and the mean of the distribution 
coincide in the large-\!$N$ limit. Both assumptions have been proven for the case of the real symmetric
random tensor \cite{parisi,example,randommat,secondmoment}, but not for the complex random tensor to our knowledge. 
We hope future study will support our results by proving these assumptions. 

This paper has added another example which shows the powerfulness of the quantum field theoretical approach
to the eigenvalue/vector distribution. We hope we will extend it further to other various cases 
which have not yet been covered.  It is also interesting to extend it for the quantities other than means.
Computing the second moment is particularly interesting, since this proves the latter assumption 
above \cite{secondmoment}.

\section*{Acknowledgment}
The author is supported in part by JSPS KAKENHI Grant No.19K03825. 
He would like to thank N.~Delporte, O.~Evnin, and L.~Lionni for some stimulating discussions.
He is especially thankful to L.~Lionni for communicating with G.~Aubrun and C.~Lancien,
whom he would like to thank for introducing him some related results and references. He would also like
to thank S.~Dartois for informing him of their upcoming paper.


%

\vspace{0.2cm}
\noindent

\let\doi\relax


\appendix

\section{Complex conventions}
\label{app:conv}
For a complex variable $x$, we define the integration measure to be
\[
dx:=dx_R dx_I,
\]
where $x_R$ and $x_I$ are the real and the imaginary parts of $x=x_R+I x_I$, respectively. 

We have a formula,
\s[
\int d\lambda  \, e^{I\lambda f^*+I \lambda^* f}
&=\int_{\mathbb{R}^2} d\lambda_R d\lambda_I  \, e^{2 I (\lambda_R f_R+\lambda_I f_I)}\\
&=\pi^2 \delta(f_R) \delta(f_I),
\label{eq:deltadef}
\s]
namely,
\[
\delta(f_R)\delta(f_I)=\frac{1}{\pi^2} \int d\lambda  \, e^{I\lambda f^*+I \lambda^* f}.
\]

The Gaussian integration for $a>0$ is
\[
\int dx \, e^{-a\, x^* x} =\int_{\mathbb{R}^2} dx_R dx_I\, e^{ -a (x_R^2+x_I^2)}=\frac{\pi}{a}.
\]
Similarly for a positive-definite $\bar N\times \bar N$ matrix $A$, 
\[
\int dx\, e^{-x^* A x}=\frac{\pi^{\bar N}}{\det A}.
\]

In this paper, we have four kinds of fermions $\bpsi_{(i)a},\psi_{(i)}^a,\bvph_{(i)}^a,\vph_{(i)a}\ (i=1,2,3,\ a=1,2,\ldots,N_i)$.
The integration measure of the fermions is denoted by $d\bar \psi d\psi d\bvph d\vph$, and  
the overall sign and the normalization are taken so that
\[
\int d\bar \psi d\psi d\bvph d\vph \, e^{\bpsi_{(i)a}\psi_{(i)}^a+\bvph_{(i)}^a\vph_{(i)a}}=1,
\label{eq:fermionintdef}
\]
where the repeated indices are assumed to be summed over. 

By explicit computation, one finds
\[
\det \left(
\begin{array}{cc}
\frac{\partial f}{\partial v} & \frac{\partial f^*}{\partial v}  \\
\frac{\partial f}{\partial v^*} & \frac{\partial f^*}{\partial v^*} 
\end{array}
\right)
=
\det \left(
\begin{array}{cc}
\frac{\partial f_R}{\partial v_R} & \frac{\partial f_I}{\partial v_R}  \\
\frac{\partial f_R}{\partial v_I} & \frac{\partial f_I}{\partial v_I} 
\end{array}
\right).
\]

\section{Integration over $C$ and $\lambda$}
\label{app:comp}
In this appendix, we will show the details of the integration over $C$ and $\lambda$.

\subsection{Integration over $C$}
\label{app:intc}
Let us first integrate over $C$. The terms containing $C$ in \eq{eq:bareaction} can be rewritten as
\s[
S_C&=-\alpha\, C^*\cdot C \\
&\hspace{1cm}-\left( C^* \cdot \bpsi_{(i)} \vph_{(j)} v_{(k)}+C\cdot \bvph_{(i)} \psi_{(j)} v_{(k)}^*
+\frac{I}{2} C \cdot \lambda^*_{(i)}v_{(j)}^* v_{(k)}^*+\frac{I}{2} C^* \cdot \lambda_{(i)} v_{(j)} v_{(k)} \right) \\
&=-\alpha \left( C^* +\frac{1}{\alpha} \left(\bvph_{(i)} \psi_{(j)} v_{(k)}^*+\frac{I}{2} \lambda^*_{(i)}v_{(j)}^* v_{(k)}^*\right)
\right)\cdot
\left(C+\frac{1}{\alpha}\left( \bpsi_{(i')} \vph_{(j')} v_{(k')}+\frac{I}{2}\lambda_{(i')} v_{(j')} v_{(k')}\right)
\right)\\
&\hspace{1cm}+\frac{1}{\alpha} \left(\bvph_{(i)} \psi_{(j)} v_{(k)}^*+\frac{I}{2} \lambda^*_{(i)}v_{(j)}^* v_{(k)}^*\right)\cdot
\left( \bpsi_{(i')} \vph_{(j')} v_{(k')}+\frac{I}{2}\lambda_{(i')} v_{(j')} v_{(k')}\right),
\label{eq:sc}
\s]
where $(i,j,k)$ and $(i',j',k')$ are summed over all the permutations of $(1,2,3)$ as in \eq{eq:bareaction}. 
We will use this rather abusive notation of neglecting the summation symbol for simplicity, unless it causes confusion.
Since $\int_{\mathbb{R}^2} dx dy\, e^{-(x+Iy + a )(x-Iy+b)}=\pi$ for any complex numbers $a,b$, we 
obtain
\[
\frac{1}{A}\int dC\, e^{S_C}=\exp\left [ \frac{1}{\alpha} \left(\bvph_{(i)} \psi_{(j)} v_{(k)}^*+\frac{I}{2} \lambda^*_{(i)}v_{(j)}^* v_{(k)}^*\right)\cdot
\left( \bpsi_{(i')} \vph_{(j')} v_{(k')}+\frac{I}{2}\lambda_{(i')} v_{(j')} v_{(k')}\right)\right].
\]
Then the action after the $C$ integration is given by
\s[
S_1=&\sum_{i=1}^3 \left(  \bpsi_{(i)}\cdot \psi_{(i)}+ \bvph_{(i)}\cdot \vph_{(i)}
+ I \lambda^*_{(i)} \cdot v_{(i)} +I \lambda_{(i)}\cdot v^*_{(i)}  \right)\\
& +\frac{1}{\alpha} \left(\bvph_{(i)} \psi_{(j)} v_{(k)}^*+\frac{I}{2} \lambda^*_{(i)}v_{(j)}^* v_{(k)}^*\right)\cdot
\left( \bpsi_{(i')} \vph_{(j')} v_{(k')}+\frac{I}{2}\lambda_{(i')} v_{(j')} v_{(k')}\right)+S_{\rm gauge},
\label{eq:s1}
\s]
where $S_{\rm gauge}$ denotes the terms multiplied with $\beta$ in \eq{eq:bareaction}.

\subsection{Integration over $\lambda_{(1)}$}
In the next step, we integrate over $\lambda_{(1)}$. The terms depending on $\lambda_{(1)}$ in \eq{eq:s1} 
can be rewritten as
\s[
S_{\lambda_1}=&-\frac{|v_{(2)}|^2 |v_{(3)}|^2}{\alpha} \lambda_{(1)}^*\cdot \lambda_{(1)}+
I \lambda^*_{(1)} \cdot v_{(1)} +I \lambda_{(1)}\cdot v^*_{(1)}\\
&+\frac{1}{\alpha}\lambda_{(1)}^* v_{(2)}^* v_{(3)}^* \cdot \left( 
I \bpsi_{(i)} \vph_{(j)} v_{(k)}-\lambda_{(2)} v_{(3)} v_{(1)}-\lambda_{(3)} v_{(1)} v_{(2)}
\right)\\
&+\frac{1}{\alpha}\lambda_{(1)} v_{(2)} v_{(3)} \cdot \left( 
I \bvph_{(i)} \psi_{(j)} v_{(k)}^*-\lambda_{(2)}^* v_{(3)}^* v_{(1)}^*-\lambda_{(3)}^* v_{(1)}^* v_{(2)}^*
\right)\\
&\hspace{-.5cm}=
-\frac{|v_{(2)}|^2 |v_{(3)}|^2}{\alpha} \left( \lambda_{(1)}^*+\cdots\right) \cdot \left( \lambda_{(1)}+\cdots\right) \\
&
-\frac{\alpha |v_{(1)}|^2}{|v_{(2)}|^2 | v_{(3)}|^2} -\frac{1}{|v_{(2)}|^2 |v_{(3)}|^2}
\left( 
v_{(1)}v_{(2)}v_{(3)} \cdot \bvph_{(i)} \psi_{(j)} v_{(k)}^*+v_{(1)}^*v_{(2)}^*v_{(3)}^* \cdot \bpsi_{(i)} \vph_{(j)} v_{(k)}
\right) \\
&-\frac{1}{\alpha |v_{(2)}|^2 |v_{(3)}|^2} \left( v_{(2)}^* v_{(3)}^* \cdot \bpsi_{(i)} \vph_{(j)} v_{(k)}\right)
\cdot \left( v_{(2)} v_{(3)} \cdot \bvph_{(i')} \psi_{(j')} v_{(k')}^*\right) \\
&-\frac{I |v_{(1)}|^2}{|v_{(2)}|^2 |v_{(3)}|^2} \left(
|v_{(3)}|^2 \left( \lambda_{(2)}^* \cdot v_{(2)}+\lambda_{(2)}\cdot v^*_{(2)}\right)+|v_{(2)}|^2 \left( \lambda_{(3)}^* \cdot v_{(3)}+\lambda_{(3)}\cdot v^*_{(3)}\right)
\right)\\
&-\frac{I}{\alpha |v_{(2)}|^2 |v_{(3)}|^2} v_{(1)} v_{(2)} v_{(3)} \cdot  \bvph_{(i)} \psi_{(j)} v_{(k)}^* \left(
|v_{(3)}|^2 \lambda_{(2)} \cdot v_{(2)}^* +|v_{(2)}|^2 \lambda_{(3)} \cdot v_{(3)}^* 
\right)\\
&-\frac{I}{\alpha |v_{(2)}|^2 |v_{(3)}|^2} v^*_{(1)} v^*_{(2)} v^*_{(3)} \cdot \bpsi_{(i)} \vph_{(j)} v_{(k)}  \left(
|v_{(3)}|^2 \lambda_{(2)}^* \cdot v_{(2)} +|v_{(2)}|^2 \lambda_{(3)}^* \cdot v_{(3)} 
\right)\\
&+\frac{|v_{(1)}|^2}{\alpha |v_{(2)}|^2 |v_{(3)}|^2}  \left(
|v_{(3)}|^2 \lambda_{(2)} \cdot v_{(2)}^* +|v_{(2)}|^2 \lambda_{(3)} \cdot v_{(3)}^* 
\right) \left(
|v_{(3)}|^2 \lambda_{(2)}^* \cdot v_{(2)} +|v_{(2)}|^2 \lambda_{(3)}^* \cdot v_{(3)} \right)
\label{eq:sintlam1}
\s]
Therefore the integral over $\lambda_1$ generates the overall factor of $(\pi \alpha /(|v_{(2)}|^2 |v_{(3)}|^2))^{N_1}$,
 and also generates the terms on the last six lines, which are added to the action.
 
 \subsection{Integrations over $\lambda_{(2)},\lambda_{(3)}$}
 Let us first extract the quadratic terms of $\lambda_{(2)},\lambda_{(3)}$. These come from the last term of \eq{eq:sintlam1}
 and such terms in \eq{eq:s1}.  Adding these, we find
 \s[
 &\frac{|v_{(1)}|^2}{\alpha |v_{(2)}|^2 |v_{(3)}|^2}  \left(
|v_{(3)}|^2 \lambda_{(2)} \cdot v_{(2)}^* +|v_{(2)}|^2 \lambda_{(3)} \cdot v_{(3)}^* 
\right) \left(
|v_{(3)}|^2 \lambda_{(2)}^* \cdot v_{(2)} +|v_{(2)}|^2 \lambda_{(3)}^* \cdot v_{(3)} \right) \\
&-\frac{|v_{(1)}|^2}{\alpha} \left( \lambda_{(2)}^* v_{(3)}^*+\lambda_{(3)}^* v_{(2)}^*\right)\cdot
\left( \lambda_{(2)} v_{(3)}+\lambda_{(3)} v_{(2)}\right)\\
&= -\frac{|v_{(1)}|^2 |v_{(3)}|^2}{\alpha} \lambda_{(2)}^*\cdot  I_{(2)}^\perp\cdot \lambda_{(2)} 
-\frac{|v_{(1)}|^2 |v_{(2)}|^2}{\alpha} \lambda_{(3)}^*\cdot  I_{(3)}^\perp\cdot \lambda_{(3)}, 
\label{eq:perpkin}
 \s]
 where $I_{(i)}^\perp$ is defined in \eq{eq:iperp}. Therefore, only the transverse components have
 quadratic terms. 
 
 As for the linear terms of $\lambda_{(2)},\lambda_{(3)}$, 
 let us first take such terms on the second line of \eq{eq:s1} and the terms 
 on the fifth and the sixth lines in \eq{eq:sintlam1}. We obtain
 \s[
&\frac{I}{\alpha} \left(\bpsi_{(i)} \vph_{(j)} v_{(k)}\cdot \left( \lambda^*_{(2)}v_{(1)}^* v_{(3)}^* + \lambda^*_{(3)}v_{(1)}^* v_{(2)}^*\right)+ \bvph_{(i)} \psi_{(j)} v_{(k)}^* \cdot \left( \lambda_{(2)} v_{(1)} v_{(3)}+ \lambda_{(3)} v_{(1)} v_{(2)}\right)\right) \\
&-\frac{I}{\alpha |v_{(2)}|^2 |v_{(3)}|^2} v_{(1)} v_{(2)} v_{(3)} \cdot  \bvph_{(i)} \psi_{(j)} v_{(k)}^* \left(
|v_{(3)}|^2 \lambda_{(2)} \cdot v_{(2)}^* +|v_{(2)}|^2 \lambda_{(3)} \cdot v_{(3)}^* 
\right)\\
&-\frac{I}{\alpha |v_{(2)}|^2 |v_{(3)}|^2} v^*_{(1)} v^*_{(2)} v^*_{(3)} \cdot \bpsi_{(i)} \vph_{(j)} v_{(k)}  \left(
|v_{(3)}|^2 \lambda_{(2)}^* \cdot v_{(2)} +|v_{(2)}|^2 \lambda_{(3)}^* \cdot v_{(3)} 
\right)\\
&=
\frac{I}{\alpha}\left(\bpsi_{(i)} \vph_{(j)} v_{(k)}\cdot \left( \lambda^{\perp*}_{(2)}v_{(1)}^* v_{(3)}^* 
+ \lambda^{\perp*}_{(3)}v_{(1)}^* v_{(2)}^*\right)+ \bvph_{(i)} \psi_{(j)} v_{(k)}^* \cdot 
\left( \lambda_{(2)}^\perp v_{(1)} v_{(3)}+ \lambda_{(3)}^\perp v_{(1)} v_{(2)}\right)\right),
\label{eq:perplin}
\s]
where $\lambda_{(i)}^\perp=I_{(i)}^\perp \lambda_{(i)}$. 

Next we take all the remaining terms linear in $\lambda_{(2)},\lambda_{(3)}$, which are on the first line of \eq{eq:s1}, 
on the fourth line of \eq{eq:sintlam1}, and in the gauge fixing terms (the last terms in \eq{eq:bareaction}). We obtain
\s[
&I \sum_{i=2,3} \left( \lambda^*_{(i)} \cdot v_{(i)} + \lambda_{(i)}\cdot v^*_{(i)} \right)\\
&-\frac{I |v_{(1)}|^2}{|v_{(2)}|^2 |v_{(3)}|^2} \left(
|v_{(3)}|^2 \left( \lambda_{(2)}^* \cdot v_{(2)}+\lambda_{(2)}\cdot v^*_{(2)}\right)+|v_{(2)}|^2 \left( \lambda_{(3)}^* \cdot v_{(3)}+\lambda_{(3)}\cdot v^*_{(3)}\right) 
\right)\\
&
+I \beta \sum_{i=2,3} ( \lambda_{(i)}^*\cdot v_{(i)}- \lambda_{(i)}\cdot v_{(i)}^*)( v_{(i)}\cdot v_{(i)} -  v_{(i)}^*\cdot v_{(i)} ^*)\\
&=
I \sum_{i=2,3} \Bigg( \lambda_{(i)}^{\parallel*} |v_{(i)}| \left( -\frac{|v_{(1)}|^2}{|v_{(i)}|^2} +1+\beta ( v_{(i)}\cdot v_{(i)} -  v_{(i)}^*\cdot v_{(i)} ^*) \right) \\
&\hspace{2cm} +\lambda_{(i)}^{\parallel} |v_{(i)}| \left( -\frac{|v_{(1)}|^2}{|v_{(i)}|^2} +1-\beta ( v_{(i)}\cdot v_{(i)} -  v_{(i)}^*\cdot v_{(i)} ^*) \right)
\Bigg),
\label{eq:lampar}
\s]
where $\lambda_{(i)}^\parallel=v_{(i)}^*\cdot \lambda_{(i)}/|v_{(i)}|$.

As we have seen above, the parallel components $\lambda_{(2)}^\parallel,\lambda_{(3)}^\parallel$ 
appear only in the linear terms in \eq{eq:lampar}.
Integrating over them results in an overall factor (see \eq{eq:deltadef}),
\[
\frac{\pi^4}{\beta^2} \delta^4(\cdots),
\]
where $\delta^4(\cdots)$ is defined in \eq{eq:deltafn}.

The transverse components $\lambda_{(2)}^\perp,\lambda_{(3)}^\perp$ exist in \eq{eq:perpkin} and \eq{eq:perplin}.
Integrating over them generates four-fermi interaction terms. More explicitly, we obtain
\s[
&-\frac{|v_{(1)}|^2 |v_{(3)}|^2}{\alpha} \lambda_{(2)}^*\cdot  I_{(2)}^\perp\cdot \lambda_{(2)} 
-\frac{|v_{(1)}|^2 |v_{(2)}|^2}{\alpha} \lambda_{(3)}^*\cdot  I_{(3)}^\perp\cdot \lambda_{(3)} \\
&+\frac{I}{\alpha}\left(\bpsi_{(i)} \vph_{(j)} v_{(k)}\cdot \left( \lambda^{\perp*}_{(2)}v_{(1)}^* v_{(3)}^* 
+ \lambda^{\perp*}_{(3)}v_{(1)}^* v_{(2)}^*\right)+ \bvph_{(i)} \psi_{(j)} v_{(k)}^* \cdot 
\left( \lambda_{(2)}^\perp v_{(1)} v_{(3)}+ \lambda_{(3)}^\perp v_{(1)} v_{(2)}\right)\right)\\
&=-\frac{|v_{(1)}|^2 |v_{(3)}|^2}{\alpha} 
\left( \lambda_{(2)}^{\perp*} -\frac{I}{|v_{(1)}|^2 |v_{(3)}|^2} I_{(2)}^{\perp *} v_{(1)} v_{(3)}\cdot \bvph_{(i)} \psi_{(j)} v^*_{(k)} \right) \\
&\hspace{5cm} 
\cdot \left( \lambda_{(2)}^{\perp} -\frac{I}{|v_{(1)}|^2 |v_{(3)}|^2} I_{(2)} ^\perp v^*_{(1)} v^*_{(3)}\cdot \bpsi_{(i)} \vph_{(j)} v_{(k)} \right)\\
&-\frac{|v_{(1)}|^2 |v_{(2)}|^2}{\alpha} 
\left( \lambda_{(3)}^{*\perp} -\frac{I}{|v_{(1)}|^2 |v_{(2)}|^2} I_{(3)} ^\perp v_{(1)} v_{(2)}\cdot \bvph_{(i)} \psi_{(j)} v^*_{(k)} \right) \\
&\hspace{5cm} 
\cdot \left( \lambda_{(3)}^{\perp} -\frac{I}{|v_{(1)}|^2 |v_{(2)}|^2} I_{(3)} ^\perp v^*_{(1)} v^*_{(2)}\cdot \bpsi_{(i')} \vph_{(j')} v_{(k')} \right)\\
&-\frac{1}{\alpha |v_{(1)}|^2 |v_{(3)}|^2}  \left( v_{(1)} v_{(3)} \cdot \bvph_{(i)} \psi_{(j)} v^*_{(k)} \right) \cdot
I_{(2)}^\perp \cdot \left( v^*_{(1)} v^*_{(3)} \cdot \bpsi_{(i')} \vph_{(j')} v_{(k')} \right)\\
&-\frac{1}{\alpha |v_{(1)}|^2 |v_{(2)}|^2}  \left( v_{(1)} v_{(2)} \cdot \bvph_{(i)} \psi_{(j)} v^*_{(k)} \right) \cdot
I_{(3)}^\perp \cdot \left( v^*_{(1)} v^*_{(2)} \cdot \bpsi_{(i')} \vph_{(j')} v_{(k')} \right).
\label{eq:lamperpint}
\s]
Therefore integrating over $\lambda_{(2)}^\perp,\lambda_{(3)}^\perp$ results in an overall factor 
of $(\pi \alpha/(|v_{(1)}|^2 |v_{(3)})|^2)^{N_2-1}$ and $(\pi \alpha/(|v_{(1)}|^2 |v_{(2)}|^2))^{N_3-1}$, respectively.
By introducing the projectors to the parallel directions to $v_{(i)}$ by 
\[
(I_{(i)}^\parallel)_{a}{}^{b}=\frac{v_{(i)a} v_{(i)}^{*b}}{|v|^2},
\]
these terms can be rewritten as
\[
-\frac{1}{\alpha}  \bvph_{(i)} \psi_{(j)} v^*_{(k)} \cdot \left(I_{(1)}^\parallel I_{(2)}^\parallel I_{(3)}^\perp+
I_{(1)}^\parallel I_{(2)}^\perp I_{(3)}^\parallel\right)\cdot \bpsi_{(i')} \vph_{(j')} v_{(k')}.
\]

This four-fermi term is added to the other four-fermi terms, which exist on the second line of \eq{eq:s1} and on the 
third line of \eq{eq:sintlam1}. We obtain 
\s[
&\frac{1}{\alpha} \bvph_{(i)} \psi_{(j)} v_{(k)}^*\cdot \bpsi_{(i')} \vph_{(j')} v_{(k')}
-\frac{1}{\alpha |v_{(2)}|^2 |v_{(3)}|^2} \left( v_{(2)}^* v_{(3)}^* \cdot \bpsi_{(i)} \vph_{(j)} v_{(k)}\right)
\cdot \left( v_{(2)} v_{(3)} \cdot \bvph_{(i')} \psi_{(j')} v_{(k')}^*\right) \\
&-\frac{1}{\alpha}  \bvph_{(i)} \psi_{(j)} v^*_{(k)} \cdot \left(I_{(1)}^\parallel I_{(2)}^\parallel I_{(3)}^\perp+
I_{(1)}^\parallel I_{(2)}^\perp I_{(3)}^\parallel\right)\cdot \bpsi_{(i')} \vph_{(j')} v_{(k')}\\
&= \frac{1}{\alpha} \bvph_{(i)} \psi_{(j)} v_{(k)}^*\cdot 
\Big[
(I_{(1)}^\parallel+I_{(1)}^\perp)(I_{(2)}^\parallel+I_{(2)}^\perp)(I_{(3)}^\parallel+I_{(3)}^\perp)
-(I_{(1)}^\parallel+I_{(1)}^\perp)I_{(2)}^\parallel I_{(3)}^\parallel \\
&\hspace{5cm}
-I_{(1)}^\parallel I_{(2)}^\parallel I_{(3)}^\perp
-I_{(1)}^\parallel I_{(2)}^\perp I_{(3)}^\parallel
\Big]
\cdot \bpsi_{(i')} \vph_{(j')} v_{(k')}
\\
&=\frac{1}{\alpha} \bvph_{(i)} \psi_{(j)} v_{(k)}^*\cdot 
\Big[
I_{(1)}^\perp I_{(2)}^\perp I_{(3)}^\parallel+I_{(1)}^\perp I_{(2)}^\parallel I_{(3)}^\perp
+I_{(1)}^\parallel I_{(2)}^\perp I_{(3)}^\perp+I_{(1)}^\perp I_{(2)}^\perp I_{(3)}^\perp
\Big]\cdot \bpsi_{(i')} \vph_{(j')} v_{(k')}
\\
&=\sum_{(ijk)\in\{ (123)\}}
\frac{|v_{(k)}|^2}{2 \alpha} \left( \bvph_{(i)} \psi_{(j)}+\bvph_{(j)} \psi_{(i)}\right) \cdot I_{(i)}^\perp I_{(j)}^\perp\cdot
\left( \bpsi_{(i)} \vph_{(j)} + \bpsi_{(j)} \vph_{(i)} \right),
\label{eq:fourfermiterm}
\s]
where we have used $I^\parallel_{(i)}\cdot v_{(i)}=v_{(i)},\ I^\perp_{(i)} \cdot v_{(i)}=0$ to obtain the last line. 

Now we collect all the computations of this appendix section. 
The overall factor of \eq{eq:rhosfourfermi} can be obtained by multiplying
all the factors generated in the integrating processes. In particular, $\delta^4(\cdots)$
comes from the integrations over $\lambda_{(2)}^\parallel,\lambda_{(3)}^\parallel$.
We also include the first term on the second line of \eq{eq:sintlam1}
into the factor.
We obtain the four-fermi action \eq{eq:sfourfermi} 
by collecting the first two terms of \eq{eq:s1}, those on the second line of \eq{eq:sintlam1}, 
those on the third line of \eq{eq:bareaction}, and \eq{eq:fourfermiterm}. 
Here we have put $|v_{(i)}|=|v|$ because of the delta functions $\delta^4(\cdots)$. 

\section{An identity}
\label{app:identity}
In this appendix we will prove an identity used in the text. Consider
\[
Z\left(\{N_i\}\right)=\exp \left( \frac{\partial}{\partial x}G\frac{\partial}{\partial x} \right) \prod_{m=1}^M 
(x H_m x)^{N_m},
\]
where $G,H_m \ (m=1,2,\ldots,M)$ are symmetric matrices of size $R\times R$, $x$ is an $R$-dimensional vector,
$N_m\ (m=1,2,\ldots,M)$ are non-negative integers. Here indices of vectors and matrices
are suppressed for notational simplicity, such as
\s[
&\frac{\partial}{\partial x}G\frac{\partial}{\partial x}=\frac{\partial}{\partial x_i}G_{ij} \frac{\partial}{\partial x_j},\\
&x H_m x=x_i (H_m)^{ij} x_j,
\s]
where contracted upper and lower indices are assumed to be summed over. 
Then consider a generating function, 
\[
Q\left(\{l_i\}\right)=\sum_{N_1,N_2,\cdots,N_M=0}^\infty Z\left(\{N_i\}\right) \prod_{m=1}^M \frac{1}{N_m!} l_m^{N_m}
\] 
with auxiliary expansion parameters $l_m\ (m=1,2,\ldots,M)$. Then $Q\left(\{l_i\}\right)$ is given by
\[
Q\left(\{l_i\}\right)=\left( \det\left(1-4 \tilde H G\right)\right)^{-\frac{1}{2}} \exp
\left( 
x\left(1-4 \tilde H G \right)^{-1} \tilde H x
\right),
\label{eq:qidentity}
\]
where $\tilde H =\sum_{m=1}^M  l_m H_{m}$. Hence $Z\left(\{N_i\}\right)$ can be computed by expanding 
\eq{eq:qidentity} in $l_m$.

To prove that, we first observe that
\[
Q\left(\{l_i\}\right)=\exp \left(\frac{\partial}{\partial x}G\frac{\partial}{\partial x} \right) \exp\left( x \tilde H x \right).
\label{eq:qkm}
\]
Since $Z\left(\{N_i\}\right)$ is obviously a polynomial function of $G$ and $H_m$
of a finite order, it is enough to prove the statement under the assumption
that $\tilde H$ is negative-definite and $G$ is positive-definite, and then consider the statement to hold generally. 
Under the assumption, we can rewrite
\[
\exp \left( x \tilde H x  \right)=\frac{1}{(2 \pi)^R} \int_{\mathbb{R}^{2 R}} d\lambda dy\, e^{y \tilde H y+I \lambda (x-y) },
\]
since $\int_{\mathbb{R}^R} d\lambda\, e^{I \lambda (x-y)}=(2\pi)^R \delta^R(x-y)$.
Putting this expression into \eq{eq:qkm}, we obtain
\[
Q\left(\{l_i\}\right)=\frac{1}{(2\pi)^R}  \int_{\mathbb{R}^{2 R}} d\lambda dy\, e^{y \tilde H y+I \lambda (x-y)-\lambda G \lambda }.
\]
This indeed derives $\eq{eq:qidentity}$ by integrating over $\lambda,y$.

\section{Derivation of \eq{eq:signrelation}}
\label{app:sign}
In this appendix we will prove \eq{eq:signrelation}, which is the key equation to obtain the gauge invariant
distribution. It should be possible to prove it by an explicit computation, but we can circumvent it by 
the following argument. Since the final result  \eq{eq:semifinal} does not depend on 
$\beta$ ($\beta^2$ which appeared in \eq{eq:rhosfourfermi} after the bosonic integrations is canceled by 
the fermionic integration \eq{eq:fermiint12}),
it is enough to prove \eq{eq:signrelation} in the lowest order of $\beta$. 

For $\beta=0$, the matrix $M(v,C,\beta=0)$ has two zero eigenvalues corresponding to the two-dimensional
degeneracy \eq{eq:2dimu1}. 
Since $M(v,C,\beta=0)$ is an hermitian matrix, one can take an orthonormal coordinates to represent the matrix
in the form,
\[
M(v,C,\beta=0)=\left( 
\begin{array}{cc}
0 & 0 \\
0 & M_{\neq 0}
\end{array}
\right),
\]
where the 2 by 2 matrix at the left upper corner vanishes because of the two zero eigenvalues,
and $M_{\neq 0}$ is an hermitian matrix with non-zero eigenvalues.
Then for $\beta\neq 0$, it has the form
\[
M(v,C,\beta)=\left( 
\begin{array}{cc}
M_0 & M_1 \\
M_2  & M_{\neq 0}+M_3
\end{array}
\right),
\]
where $M_i\propto \beta$. In particular, $M_0$ is a matrix with components $(M_0)_{ij}=e_i^\dagger M(v,C,\beta) e_j$,
where $e_i \ (i=1,2)$ are the orthonormalized eigenvectors of the zero eigenvalues of $M(v,C,\beta=0)$.  
It is obvious that 
\[
\hbox{The lowest order in $\beta$ of}\det M(v,C,\beta)=\det M_0 \det M_{\neq 0}.
\label{eq:lowest}
\]

To compute $\det M_0$, let us first fix the convention. 
The matrix $M(v,C,\beta)$ is the matrix which represents the fermionic (quadratic) terms
of $S$ in \eq{eq:bareaction}:
\[
\hbox{Fermion terms of }S=\bar \Psi  M(v,C,\beta) \Psi,
\]
where $\bar \Psi=\left( \bpsi_{(1)}\ \bpsi_{(2)}\ \bpsi_{(3)}\ \bvph_{(1)}\ \bvph_{(2)}\ \bvph_{(3)}\right)$ and 
$\Psi=\left(\psi_{(1)}\ \psi_{(2)}\ \psi_{(3)}\ \vph_{(1)}\ \vph_{(2)}\ \vph_{(3)} \right)$. 
For $\beta=0$, we have
\[
M(v,C,\beta=0)=\left( 
\begin{array}{cccccc}
\text{Id} &0 &0 &0 &-C^*\cdot v_{(3)} & -C^*\cdot v_{(2)} \\
0 &\text{Id} &0 &-C^*\cdot  v_{(3)} & 0&-C^*\cdot v_{(1)} \\
0&0&\text{Id} &-C^*\cdot v_{(2)} & -C^*\cdot v_{(1)}&0\\
0 &-C\cdot v_{(3)}^* & -C\cdot v_{(2)}^*&\text{Id}  &0 &0  \\
 -C\cdot  v_{(3)}^* & 0&-C\cdot v_{(1)}^* &0 &\text{Id}  &0\\
-C\cdot v_{(2)}^* & -C\cdot v_{(1)}^*&0&0&0&\text{Id} 
\end{array}
\right),
\]
where $\text{Id}$ denotes the identity matrix.
It is straightforward to find that the eigenvectors of the zero eigenvalues of $M(v,C,\beta=0)$ are given by
\[
\Psi(\alpha)=(\alpha_1  v_{(1)}^*, \alpha_2  v_{(2)}^*,\alpha_3  v_{(3)}^*,
-\alpha_1  v_{(1)}, -\alpha_2  v_{(2)},-\alpha_3  v_{(3)})
\] 
with $\sum_{i=1}^3 \alpha_i=0$. This can be checked by explicitly computing $M(v,C,\beta=0)\Psi(\alpha)$.
For instance, the first component is
\s[
\left( M(v,C,\beta=0)\Psi(\alpha)\right)_1&=\alpha_1 v_{(1)}^* +\alpha_2 C^*\cdot v_{(3)} v_{(2)} +\alpha_3 C^*\cdot
v_{(2)} v_{(3)}\\&=(\alpha_1+\alpha_2+\alpha_3) v_{(1)}^*\\&=0,
\s]
where we have used the eigenvector equations \eq{eq:egeqs}. Therefore we take an orthonormal coordinate 
of the two-dimensional zero eigenvalue subspace in terms of $\Psi(\alpha)$ with 
\[
( \alpha_1,\alpha_2,\alpha_3)=\frac{1}{2 \sqrt{3} |v|} (-2,1,1),\ \frac{1}{2 |v|} (0,1,-1).
\label{eq:egal}
\]

The $\beta$ linear term of $M(v,C,\beta)$ is given by
\s[
2 \beta 
\left( 
\begin{array}{cccccc}
0 &0 &0 &0 &0&0\\
0 &v_{(2)} v_{(2)} &0 &0 &- v_{(2)} v_{(2)}^* &0  \\
0& 0 &v_{(3)} v_{(3)}  &0 &0&- v_{(3)} v_{(3)}^* \\
0 &0 &0 &0 &0&0\\
0 &-v_{(2)}^* v_{(2)} &0 &0 & v_{(2)}^* v_{(2)}^* &0  \\
0& 0 &-v_{(3)}^* v_{(3)}  &0 &0&v_{(3)}^* v_{(3)}^* 
\end{array}
\right).
\s]
Then the matrix $M_0$ can explicitly be computed by sandwiching this matrix by the eigenvectors $\Psi(\alpha)$ 
with \eq{eq:egal} and 
the conjugates. We obtain
\[
M_0=2 \beta \left(
\begin{array}{cc}
\frac{1}{3}(v_{(2)}\cdot v_{(2)}+v_{(3)}\cdot v_{(3)})& \frac{1}{\sqrt{3}} (v_{(2)}\cdot v_{(2)}-v_{(3)}\cdot v_{(3)}) \\
 \frac{1}{\sqrt{3}} (v_{(2)}\cdot v_{(2)}-v_{(3)}\cdot v_{(3)}) & v_{(2)}\cdot v_{(2)}+v_{(3)}\cdot v_{(3)} 
\end{array}
\right),
\] 
whose determinant is $\det M_0=16 \beta^2 v_{(2)}\cdot v_{(2)}\,v_{(3)}\cdot v_{(3)}/3$.
Combining with \eq{eq:lowest}, this proves \eq{eq:signrelation}.

\section{The method using the Schwinger-Dyson equation}
\label{app:methodsd}
In this appendix, to make this paper self-contained,
we explain the method using the SD equation.
The contents are rather standard, but would be useful for unfamiliar readers.

The general form of a four-fermi theory is given by
\[
S=\bar \psi_a K_{ab} \psi_b + g J_{abcd} \bar \psi_a\bar \psi_b \psi_c \psi_d,
\]
where $g$ is a coupling, $J,K$ are constants, and $J$ satisfies 
\[
J_{abcd}=-J_{bacd}=-J_{abdc},
\]
to be consistent with the anti-commutativity of the fermions.
The partition function of the system is defined by
\[
Z=\int d\bar\psi d\psi\, e^S,
\]
where the integration measure $d\bar\psi d\psi$ is defined so that \cite{zinn}
\[
\int d\bar\psi d\psi\, e^{\bar \psi_a K_{ab} \psi_b}=\det K.
\]
The system is invariant under a scaling transformation,
\[
\bar \psi_a'= s \bar \psi_a, \ \psi_a'= s^{-1} \psi_a,
\label{eq:scale}
\]
where $s$ is arbitrary.

The fermion integration vanishes when the integrand is a total derivative. So, we have an identity,
\s[
0&=\int d\bar \psi d\psi \frac{\partial}{\partial \bar \psi_a}( \bar \psi_b \, e^S) \\
&=\int d\bar \psi d\psi \left( \delta_{ab} -K_{ac} \bar \psi_b \psi_c -2  g J_{acde} \bar \psi_b \bar \psi_c \psi_d \psi_e\right)e^S.
\s]
By dividing this by $Z$, we obtain a Schwinger-Dyson equation, 
\s[
K_{ac} \langle \bar \psi_b \phi_c \rangle +2 g J_{acde} \langle \bar \psi_b  \bar \psi_c \psi_d \psi_e \rangle=\delta_{ab},
\label{eq:appsd}
\s]
where correlation functions are defined by
\[
\langle {\cal O}\rangle =\frac{1}{Z} \int d\bar\psi d\psi\, {\cal O} e^S.
\]

The four-fermion correlation function in \eq{eq:appsd} may be expressed in terms of connected 
two-fermion\footnote{Since one-fermion correlation functions of fermions vanish, any two-fermion correlation functions are connected.} 
and four-fermion correlation functions,
\[
\langle \bar \psi_b  \bar \psi_c \psi_d \psi_e \rangle=-\langle \bar \psi_b   \psi_d  \rangle \langle \bar \psi_c \psi_e \rangle
+\langle \bar \psi_b  \psi_e  \rangle \langle \bar \psi_c  \psi_d  \rangle+\langle \bar \psi_b  \bar \psi_c \psi_d \psi_e \rangle_c,
\label{eq:fourpoint}
\]
where $\langle {\cal O} \rangle_c$ denotes the connected correlation functions, and 
we have assumed that only the correlation functions with equal numbers of $\bar \psi$ and $\psi$ remain 
non-vanishing due to the symmetry \eq{eq:scale}. Now assuming that  
$\langle \bar \psi_b  \bar \psi_c \psi_d \psi_e \rangle_c$ can be neglected as higher orders, 
\eq{eq:appsd} and \eq{eq:fourpoint} lead to an equation for the two-fermion correlation functions,
\[
K_{ac} \langle \bar \psi_b \phi_c \rangle +4 g J_{acde} 
\langle \bar \psi_b  \psi_e  \rangle \langle \bar \psi_c  \psi_d  \rangle=\delta_{ab}.
\label{eq:consist1}
\]
In fact, the equation \eq{eq:consist1} can be written in another way. Let us denote $G_{ab}=\langle \bar \psi_a \psi_b \rangle$.
Then, by applying $G^{-1}$ on both sides of \eq{eq:consist1}, we obtain
\[
K_{ab} +4 g J_{acdb} G_{cd}=G^{-1}_{ba}.
\label{eq:consist2} 
\]

Once a solution of \eq{eq:consist2} for $G$ is obtained, the partition function $Z$ can be computed 
in the following manner. We first note that
\s[
\frac{d}{d g} \log Z&= J_{abcd} \langle \bar \psi_a  \bar \psi_b \psi_c \psi_d \rangle \\
&=2  J_{abcd} G_{ad} G_{bc} 
\label{eq:dzdg}
\s]
by applying the same approximation as above expressing the four-fermion correlation functions in terms of two-fermion ones. 
This first-order differential equation can be solved with the initial condition,
\[
Z(g=0)=\det K.
\]

The procedure above can be summarized as a more intuitive process.
Let us define
\s[
S_{\rm eff}&=\langle S \rangle -\log \det G \\
&=K_{ab} G_{ab}+2 g J_{abcd} G_{ad} G_{bc} - \log \det G,
\label{eq:defofseff}
\s]
where $\langle S \rangle$ in \eq{eq:defofseff} has been computed with the similar approximation above. 
In this expression, depending on the sign of $\det G$, the term $\log \det G$ may be replaced by $\log(-\det G)$. 

The stationary condition of $S_{\rm eff}$ with respect to $G$ leads to  \eq{eq:consist2}:
\s[
0&=\frac{\partial}{\partial G_{ab} }S_{\rm eff} \\
&= K_{ab} +4 g J_{acdb} G_{cd} -G^{-1}_{ba}.
\label{eq:appstationary}
\s]
Moreover, let us define 
\[
Z^0_{\rm eff}=\det K\,  e^{S^0_{\rm eff}-S_{\rm eff}^0(g=0)},
\label{eq:appzeff}
\]
where $S^0_{\rm eff}$ is the value of $S_{\rm eff}$ for the solution $G=G^0$ of \eq{eq:appstationary}, and 
$S_{\rm eff}^0(g=0)$ is that for $g=0$. Then $Z^0_{\rm eff}$ gives the partition function in this approximation.
This is because, by using the stationary condition \eq{eq:appstationary},
\[
\frac{d}{dg}\log Z^0_{\rm eff}=\frac{\partial S^0_{\rm eff}}{\partial g}  + \frac{\partial S^0_{\rm eff}}{\partial G^0_{ab} }\frac{d G^0_{ab}}{dg}=
2  J_{abcd} G^0_{ad} G^0_{bc},
\label{eq:derz}
\]
which agrees with \eq{eq:dzdg}, and $Z^0_{\rm eff}(g=0)=\det K$. 

\section{Stationary conditions of $S_{\rm eff}$}
\label{app:stationary}
By taking $\partial S_{\rm eff}/\partial Q^i_j=0$ one obtains
\s[
&-1+\sum_{j\neq i} Q^j_1 g_j-\frac{Q^i_1}{\det Q_{(i)}}=0, \\
&\sum_{j\neq i} Q^j_2 g_j-\frac{Q^i_2}{\det Q_{(i)}}=0, \\
&\sum_{j\neq i} Q^j_3 g_j-\frac{Q^i_3}{\det Q_{(i)}}=0, \\
&1+\sum_{j\neq i} Q^j_4 g_j-\frac{Q^i_4}{\det Q_{(i)}}=0,
\label{eq:Qeq}
\s]
where $g_i=N_i g$. By defining new parameters, $A_i=(\det Q_{(i)})^{-1}$ and $B_j=\sum_{i=1}^3 Q_j^ig_i$, one can 
solve \eq{eq:Qeq} as
\[
Q_1^i=\frac{B_1-1}{g_i+A_i},\ Q_2^i=\frac{B_2}{g_i+A_i},\ Q_3^i=\frac{B_3}{g_i+A_i},\ Q_4^i=\frac{B_4+1}{g_i+A_i},
\]
with consistency conditions,
\[
&\det Q_{(i)}=\frac{(-1+B_1)(1+B_4)-B_2 B_3}{(g_i+A_i)^2} =\frac{1}{A_i},
\label{eq:QA} \\
&(-1+B_1)p=B_1,\ B_2 p =B_2,\ B_3 p =B_3,\ (1+B_4)p=B_4,
\label{eq:Beqs}
\]
where
\[
p=\sum_{i=1}^3 \frac{g_i}{g_i+A_i}.
\label{eq:p}
\]
From the equations \eq{eq:Beqs}, we immediately obtain $p\neq 1$, 
and so $B_2=B_3=0$ and $B_1=-B_4=p/(p-1)$. Putting these into \eq{eq:QA}, we obtain 
\[
A_i+(p-1)^2 (g_i+A_i)^2=0.
\label{eq:A}
\]
Let us summarize: For given $g,N_i$ (note that $g_i=gN_i$), the four equations in \eq{eq:p} and \eq{eq:A} determine 
the four variables $p$ and $A_i$, and therefore all $Q_j^i$. 

The quadratic equations in \eq{eq:A} have two solutions for $A_i$. Which solution should be taken for small $g$ 
can be determined by the following argument. When $g\rightarrow 0$, the four-fermi interactions vanish, and therefore
$Q_1^i=-Q_4^i=1, Q_2^i=Q^i_3=0$. This is satisfied by the solution,
\[
A_i=\frac{-1-2 g_i (p-1)^2 -\sqrt{1+4 g_i (p-1)^2}}{2 (p-1)^2}.
\]  
Putting these into \eq{eq:p} and reparameterize $p=1-q$, we obtain \eq{eq:q}.

\section{The explicit form of $h_0$ and $\det h_2$}
\label{app:h0h2}
The computations are straightforward. We obtain
\[
h_0=\frac{(-1 + l) x}{1 + 2 l} -\log(l),
\]
and
\[
\det h_2=\frac{(1 + 6 l + 8 l^3 - 12 l^2 (-1 + x)) (1 + 3 l - 4 l^3 + 6 l^2 x)^2}{27 (-1 + l)^2 l^6 (1 + 2 l)^7}.
\]

\end{document}